\documentclass[12pt]{article}
\usepackage{latexsym}
\usepackage{amsmath,,calrsfs}
\usepackage{amsfonts}
\usepackage{amssymb}
\usepackage{amscd}
\usepackage{bbm}
\usepackage{fancybox}
\usepackage{cite}
\usepackage{amsmath,amsfonts,amsbsy}
\usepackage{pstricks,pst-node}
\usepackage[small,bf,hang]{caption2}
\usepackage{graphicx}
\usepackage{epsfig}
\usepackage{psfrag}
\usepackage{comment}

\usepackage{float}

\psset{unit=1.3cm,linewidth=.5pt,radius=.2}  

\usepackage{multirow}                     
\usepackage{float}                          
\usepackage{lscape}                         
\usepackage{bm}

\newcommand{\beq}{\begin{equation}}
\newcommand{\eeq}{\end{equation}}
\newcommand{\bi}{\begin{itemize}}
\newcommand{\ei}{\end{itemize}}
\newcommand{\bt}{\begin{tabular}}
\newcommand{\et}{\end{tabular}}
\newcommand{\bc}{\begin{center}}
\newcommand{\ec}{\end{center}}

\def\theequation{\arabic{section}.\arabic{equation}}

\newcommand{\be}{\begin{equation}}
\newcommand{\ee}{\end{equation}}
\newcommand{\bea}{\begin{eqnarray}}
\newcommand{\eea}{\end{eqnarray}}
\newcommand{\ba}{\begin{array}}
\newcommand{\ea}{\end{array}}
\newcommand{\p}[1]{(\ref{#1})}
\newcommand{\lb}[1]{\label{#1}}

\def\bbox{{\,\lower0.9pt\vbox{\hrule \hbox{\vrule height 0.2 cm
\hskip 0.2 cm \vrule height 0.2 cm}\hrule}\,}}
\newcommand{\dsl}{\pa \kern-0.5em /}

\newcommand{\nn}{\nonumber \\}

\makeatletter \@addtoreset{equation}{section} \makeatother
\renewcommand{\theequation}{\thesection.\arabic{equation}}

\def\slashchar#1{\setbox0=\hbox{$#1$}           
   \dimen0=\wd0                                 
   \setbox1=\hbox{/} \dimen1=\wd1               
   \ifdim\dimen0>\dimen1                        
      \rlap{\hbox to \dimen0{\hfil/\hfil}}      
      #1                                        
   \else                                        
      \rlap{\hbox to \dimen1{\hfil$#1$\hfil}}   
      /                                         
   \fi}

\begin{document}

\begin{titlepage}

\renewcommand{\thefootnote}{\star}

\begin{center}

\hfill  {}


{\Large \bf  Quaternion-K$\bf \ddot{a}$hler ${\cal N}=4$ Supersymmetric
\vspace{0.2cm}

Mechanics}
\vspace{0.3cm}

\vspace{1.3cm}
\renewcommand{\thefootnote}{$\star$}

{\large\bf Evgeny~Ivanov} ${}^\star$,
{\quad \large\bf Luca~Mezincescu} ${}^\ast$
 \vspace{0.5cm}

{${}^\star$ \it Bogoliubov Laboratory of Theoretical Physics, JINR,}\\
{\it 141980 Dubna, Moscow region, Russia} \\
\vspace{0.1cm}

{\tt eivanov@theor.jinr.ru}\\
\vspace{0.4cm}

{${}^\ast$\it Department of Physics, University of Miami,}\\
{\it P.O. Box 248046, Coral Gables,
FL 33124, USA}\\
\vspace{0.1cm}

{\tt  mezincescu@physics.miami.edu}\\

\end{center}
\vspace{0.2cm} \vskip 0.6truecm \nopagebreak

\begin{abstract}
\noindent Using the ${\cal N}=4, 1D$ harmonic superspace approach, we construct a new type of ${\cal N}=4$ supersymmetric
mechanics involving $4n$-dimensional Quaternion-K\"ahler (QK) $1D$ sigma models as the bosonic core. The basic ingredients of our construction
are {\it local} ${\cal N}=4, 1D$ supersymmetry realized by the appropriate transformations in $1D$ harmonic superspace, the general
${\cal N}=4, 1D$ superfield vielbein and a set of $2(n+1)$ analytic ``matter'' superfields representing $(n+1)$ off-shell supermultiplets $({\bf 4, 4, 0})$.
Both superfield and component actions are given for the simplest QK models with the manifolds
$\mathbb{H}{\rm H}^n = Sp(1,n)/[Sp(1)\times Sp(n)]$ and $\mathbb{H}{\rm P}^n = Sp(1+n)/[Sp(1)\times Sp(n)]$ as the bosonic targets.
For the general case the relevant superfield action and constraints on the $({\bf 4, 4, 0})$ ``matter'' superfields are presented.
Further generalizations are briefly discussed.
\end{abstract}
\vskip 1cm

\vskip 0.5cm

\vspace{1cm}

\phantom{Keywords: }

\newpage

\end{titlepage}

\setcounter{footnote}{0}

\newpage
\setcounter{page}{1}

\section{Introduction}
Supersymmetric sigma models in one dimension amount to various versions of Supersymmetric Quantum Mechanics (SQM) \cite{Witt}.
They reveal interesting geometric properties. Sometimes they can be obtained by a direct reduction from higher-dimensional
supersymmetric theories and so inherit the target geometries associated with the latter. But there exist a wide variety of SQM models
which {\it cannot} be related in this way to any their higher-dimension analogs. They surprisingly feature the specific geometries inherent
just to the one-dimensional case \cite{cole,Hull,GPS}.

A notorious example is hyper-K\"ahler (HK) sigma models. HK manifolds are target manifolds of the ${\cal N}=2, 4D$
sigma models associated with a general self-interaction of hypermultiplets \cite{HK1,HK2}. The most general models of this kind are formulated in ${\cal N}=2, 4D$
harmonic superspace \cite{HSSa,HSSb}, where hypermultiplets are represented by analytic harmonic superfields $q^{+}(\zeta, u)$ with an infinite number of auxiliary
fields. The relevant off-shell actions were constructed in \cite{HSShk1,GIOT}. The geometric interpretation of the corresponding superfield Lagrangians as the unconstrained analytic
HK potentials was further given in \cite{HSShk2}. The direct reduction of these actions to $1D$ yields ${\cal N}=8$ supersymmetric mechanics with $4n$
dimensional HK manifolds as bosonic target spaces. However in $1D$ one can construct an ${\cal N}=4$ SQM with a generic HK target space \cite{HKsqm,DI1}
which can by no means be obtained through
the dimension reduction from  higher dimensions. This model can be also extended to the so called HKT (``hyper-K\"ahler with torsion'')
SQM models (see, e.g.,  \cite{HKT} and references therein) , and, further (for $n \geq 2$), to the ${\cal N}=4$ SQM model with
yet a more general target space geometry \cite{cole,GPS}.

When coupled to ${\cal N}=2, 4D$ supergravity, the hypermultiplet HK sigma models are deformed to those with the Quaternion-K\'ahler (QK) manifolds
as the target \cite{QK}. The deformation parameter is Einstein constant, and when it is sent to zero, the QK targets contract into the HK ones
arising as the ``flat limit'' of the former. The harmonic superspace formulation of these sigma models along the same line as in the HK case,
was given in \cite{MatN2,GBIO,GIOO,Ivanov:1999vg}.
Up to now no SQM model with the QK target geometry has been constructed. The basic reason seemingly was the difficulties with accounting
for the supergravity quantities in the quantum-mechanical context, so as to ensure, in one or another way, the original local supersymmetry and local $SU(2)$
automorphism symmetry.

The basic aim of the present paper is to fill up this gap and construct ${\cal N}=4$ supersymmetric mechanics  with an arbitrary QK bosonic target. We do it on the classical
level, leaving the case of the full-fledged QK ${\cal N}=4$ SQM for the future study. Our construction
does not refer to any kind of dimensional reduction, it is intrinsically one-dimensional. We will use
the formalism of off-shell $1D$ harmonic superspace \cite{ILe} as the most appropriate one for constructing ${\cal N}=4$, $1D$ supersymmetric sigma models \cite{DI1}.
Our starting point is a minimal local extension of the semi-direct product of rigid ${\cal N}=4, 1D$ supersymmetry and automorphism $SU(2)$,
such that it is isomorphic to the classical (having no central charge) small ${\cal N}=4$ superconformal group. Then a rather simple modification of the ${\cal N}=4$
action describing general ${\cal N}=4, 1D$ HK sigma model yields, upon fixing a gauge with respect to the time reparametrizations and local supersymmetry, the ${\cal N}=4, 1D$ sigma model
with a generic QK target manifold in the bosonic sector. We consider, in some detail, the simplest examples of ${\cal N}=4$ QK mechanics
based on the ``maximally flat'' $\mathbb{H}{\rm H}^n$ and  $\mathbb{H}{\rm P}^n$ sigma models.

We start in Section ~2 by giving some details of the bosonic Lagrangians of the $1D$ QK sigma models just mentioned. In Section~3
we introduce the basic superfields we will deal with, and define the appropriate ${\cal N}=4, 1D$ ``supergravity'',
 which is one of the basic ingredients of our construction. In Section~4 we build the invariant superfield and component
 actions for the $\mathbb{H}{\rm H}^n$ and  $\mathbb{H}{\rm P}^n$ examples. Section~5 is devoted to a generalization to the case of generic
 QK manifolds. Some further generalizations are sketched in the concluding Section~6. The Appendices A, B and C collect
 various technical details, including the full set of the local transformations of the component fields (Appendix A) and the basic
 quantities of the $\mathbb{H}{\rm H}^n$ and  $\mathbb{H}{\rm P}^n$ geometries (Appendix C).

\section{Warm-up: One-dimensional QK sigma models}
As the natural point of departure, we will start with presenting the $1D$ actions of the simplest QK sigma models associated
with the cosets $\mathbb{H}{\rm H}^n = Sp(1,n)/[Sp(1)\times Sp(n)]$ and $\mathbb{H}{\rm P}^n = Sp(1+n)/[Sp(1)\times Sp(n)]$.
In the next Section we will construct their ${\cal N}=4, 1D$ supersymmetric extensions. We will basically
follow \cite{Ivanov:1999vg} where the general action of bosonic QK sigma models, as well as its particular $\mathbb{H}{\rm H}^n$ and $\mathbb{H}{\rm P}^n$ examples, were given in
the form most convenient for our purposes.

The starting point  is the $\mathbb{H}{\rm H}^n$  action, given by formula  (2.72)  in  \cite{Ivanov:1999vg} which we dimensionally reduce so that the involved fields
have no space dependence:
\bea
S_{HH} &=&= \int dt\,L(t) = {1\over 2} \int dt \;\{\; ({\dot {\hat{F}}} {\dot {\hat{F}}}) +
\hat{\kappa}^2\;(\hat{F}_{(i r}{\dot {{\hat{F}}}}_{j)}\,^r)(\hat{F}_{s}^{(i}\,{\dot
{\hat{F}}}\,^{j)s}) \nn
&& - \;{\hat{\kappa}^2\over 2}\,{1\over 1 + {\hat{\kappa}^2\over 2}\, \hat{F}^2}
(\hat{F}{\dot {\hat{F}}})(\hat{F}{\dot {\hat{F}}}) \;\}~. \label{HP6}
\eea
Here, the  scalar fields ${\hat F}^{ir},\,  r = 1 \cdots 2n, i = 1,2$ obey the conjugation properties
\bea
\overline{({\hat F}^{ir}(x))} =
\Omega_{rs}\varepsilon_{ik}\;{\hat F}^{ks}(x)~, \label{real}
\eea
where $\Omega_{rs} = -\Omega_{sr}~, \; \Omega^{\,rs}\Omega_{\,st} =
\delta^{\,r}_{\,t}$,
is the totally antisymmetric constant $Sp(n)$ metric, and $\varepsilon_{12} = 1 = -\varepsilon_{21}~, \;
\varepsilon_{ik}\varepsilon^{kl} = \delta^l_i$, and ${\hat F}^2 = {\hat F}^{ir} {\hat  F}_{ir}$. In the process of dimensional reduction $4D \rightarrow 1D$
it was convenient for us to pass, from the original fields of the mass dimension 1 and Einstein
constant $\kappa^2$ of dimension -2, to the dimensionless fields and dimensionless $\hat{\kappa}^2$ by rescaling
$(f, F) \rightarrow \rho^{-1}(f, F), {\kappa}^2 = \rho^2 \hat{\kappa}^2, [\rho]= m^{-1}\,.$ This is allowed because
the action \p{HP6} is defined up to an overall renormalization constant.

The target space metric can be read off from \p{HP6}
\bea
\hat{g}_{ir\,js} = \varepsilon_{ij}\Omega_{rs}  -\hat{\kappa}^2 \varepsilon_{ij} \hat{F}_{kr}\hat{F}^k_s - \frac{\hat{\kappa}^2}{2}\hat{F}_{ir} \hat{F}_{js}\,
\frac{2 + \frac{\hat{\kappa}^2}{2} \hat{F}^2}{1 + \frac{\hat{\kappa}^2}{2} \hat{F}^2}\,. \lb{Metric}
\eea
The inverse metric is given by the expression
\bea
\hat{g}^{ir\,js} = \varepsilon^{ij}\Omega^{rs}  - \frac{\hat{\kappa}^2}{1 + \frac{\hat{\kappa}^2}{2} \hat{F}^2}
\varepsilon^{ij} \hat{F}_{k}^{r}\hat{F}^{ks} + \frac{\hat{\kappa}^2}{2}\hat{F}^{ir} \hat{F}^{js}\,
\frac{2 + \frac{\hat{\kappa}^2}{2} \hat{F}^2}{1 + \frac{\hat{\kappa}^2}{2} \hat{F}^2}\,. \lb{InvMetr}
\eea

With the change of variables:
\bea
F^{ir} = {1 \over\sqrt{2}\, |\hat\kappa| \omega} \;\hat{F}^{ir},\ \  \omega = {1\over \sqrt{2}\,|\hat\kappa|}\, \lb{FhatF}
\sqrt{1 + {\hat{\kappa}^2\over 2}\,\hat{F}^2},
\eea
whose inverse is
\bea
{\hat F}^{ir} = \sqrt{2}\, |\hat{\kappa}| \omega \, {F}^{ir} = \frac{F^{ir}}{\sqrt {1 -\frac{\hat{\kappa}^2}{2} F^2}},\ \
\omega = {1\over \sqrt{2}\,|\hat{\kappa}|}
\left(\frac{1}{\sqrt{1 - {\hat{\kappa}^2\over 2}\,{F}^2}}\right), \lb{hatFF}
\eea
the action \p{HP6} is rewritten as:
\bea\label{sigmamodel}
S_{HP} = {1\over 2} \int dt
\;\{\; {1\over 1-{\hat{\kappa}^2\over 2}\,F^2}({\dot F}{\dot F}) +
{\hat{\kappa}^2\over [1 - {\hat{\kappa}^2\over 2}\, F^2]^2}
(F_{ir}F^i_s)({\dot F}^r_j {\dot F}^{js})\;\}~. \label{HP7}
\eea
The action (\ref{sigmamodel}) is invariant under the isometry $Sp(1,n)$. The coset
$Sp(1,n)/[Sp(1)\times Sp(n)]$ transformations of $F^{ir}$ read
\bea
\delta F^{ir} = \lambda^{ri} - \hat{\kappa}^2 \lambda^{sj}F^r_jF^i_s\,. \lb{FtransfC}
\eea
It is easy to check that two such infinitesimal transformations close to the homogeneous ones:
\bea\label{Ftransf}
\left[\delta_2,\delta_1\right]F^{ir}= -\hat{\kappa}^2\left[(\lambda_1^{sj}\lambda_{2j}^r- \lambda_2^{sj}\lambda_{1j}^r)F^i_s +(\lambda_1^{sj}\lambda_{2s}^i
- \lambda_2^{sj}\lambda_{1s}^i)F^r_j\right ],
\eea
which just form the $Sp(1)\times Sp(n)$ subgroup of the total $Sp(1,n)$ isometry. The transformations \p{FtransfC}
were derived in \cite{GBIO}, \cite{Ivanov:1999vg}
from the appropriate $Sp(1,n)$ transformations of the harmonic superfields describing the matter
and conformal compensator hypermultiplets. This was made by fixing a gauge with respect to
some local $SU(2)$ transformations and then modifying the original constant-parameter isometry by the compensating local $SU(2)$ transformation needed to preserve this gauge.
This derivation will be repeated in Sect. 4 in the context of the considered $1D$ sigma model.

Let us verify that the  action (\ref{sigmamodel}) is invariant under the transformations in (\ref{FtransfC}). First one can show that
\bea&&\left(1-\frac{\hat{\kappa}^2}{2}F^2\right)^3\delta(2L) = \left(1-\frac{\hat{\kappa}^2}{2}F^2\right)\left[2\hat{\kappa}^2
\delta F^{ir}F_{jr}{\dot F}^{js}{\dot F}_{is} + 2\left(1-\frac{\hat{\kappa}^2}{2}F^2\right){\dot F}{\dot {\delta F}}\right. \nn
&&\left. + 2\hat{\kappa}^2 \delta F_{ir} F^i_s{\dot F}^r_j{\dot F}^{js}+ 2\hat{\kappa}^2 F_{ir} F^i_s{\dot {\delta F}}^r_j{\dot F}^{js}\right]
+ 2\hat{\kappa}^4(F\delta F) F_{ir} F^i_s{\dot {F}}^r_j{\dot F}^{js}. \lb{IdInv}
\eea
Then we have to check the cancelations of the terms of the order $\hat{\kappa}^6, \hat{\kappa}^4, \mbox{and} \,\hat{\kappa}^2$. The $\hat{\kappa}^2$
order terms can be shown to vanish by appropriately changing the summation indices, effectuating the derivatives and
using the identity:
\bea
\varepsilon_{ii^\prime}\varepsilon_{jj^\prime} = \varepsilon_{ij} \varepsilon_{i^\prime j^\prime}-\varepsilon_{i^\prime j}\varepsilon_{ij^\prime}.
\eea
In order to check the vanishing of the $\hat{\kappa}^4$ term one must compute the coefficients of the following structures:
\bea
{\dot F}^{is}{\dot F}_{js},\quad \lambda^{s^\prime j}{\dot F}^{is}{\dot F}_{j^\prime s^\prime},\quad \lambda^{s^\prime j}F_{i^\prime s^\prime }{\dot F}^{is}{\dot F}^{j^\prime r },
\eea
together with the above mentioned identity, and its consequence
\bea
A_i B_j = B_j A_i + \epsilon_{ij}A^kB_k\,,
\eea
and also the evident antisymmetry of the terms of the form: ${\dot F}^r_j{\dot F}^{js}, F_{ir}F^i_s\,$, under the interchange of $r$ and $s$.
It remains to check the $\hat{\kappa}^6$ terms. One way to do  it, is to use the relation:
\bea
F^2F_{ir}F^i_s\lambda^{s^\prime j^\prime}F_{j s^\prime}{\dot F}^r_{j^\prime}{\dot F}^{js}  = \frac{1}{2}F^2F_{ir}F^i_s\lambda^{s^\prime j^\prime}F_{j^\prime s^\prime}
{\dot F}^r_{j}{\dot F}^{js}
\eea
and, subsequently, the identity
\bea
F_{j^\prime s}\left( F_{i s^\prime}F_{r}^i\right) + F_{j^\prime s^\prime }\left(F_{ir }F_{s}^i\right) +F_{j^\prime r}\left( F_{is}F_{s^\prime}^i\right)  = 0\,,
\eea
in order to represent the corresponding variation as a trace of the product of two matrices, respectively symmetric and antisymmetric in the indices $r, s$, which is identically zero.

Thus we have proved that the right-hand side of \p{IdInv} is zero, and, hence, $\delta L =0\,$.

As a simple illustration we will consider the $Sp(1,1)/[Sp(1)\times Sp(1)]$ sigma model. In this case the index $r$
takes the values $1,2$ and we
replace it by the letter $a$: $F^{ir} \rightarrow F^{i a} ; a,i = 1,2, \, \Omega _{ab} = \varepsilon_{ab}$.
The reality condition amounts to having two complex entries.
Indeed, using the matrix notation:
\bea
F^{ia} \rightarrow F = \left(\begin{array}{cc}z^1 & z^2 \\{\bar z}^1  & {\bar z}^2\end{array}\right) : \quad \overline{({ F}^{ia})} =
\varepsilon_{a a^\prime}\varepsilon_{ii^\prime}\;{ F}^{ i^\prime a^\prime}~, \label{real1}
\eea
with the conjugation rules
\bea
{\overline {z^1}} = {\bar z}_1  = {\bar z}^2, \quad {\overline {z^2}} = {\bar z}_2  = -{\bar z}^1,
\eea
the reality condition is satisfied. In terms of these new variables (\ref{HP7}) becomes:
\bea
S_{HP} =  \int dt
\; \frac {{\dot z}^j{\dot {\bar z}}_j}{(1-\hat{\kappa}^2\,z^i{\bar z}_i)^2}. \lb{S4}
\eea
The infinitesimal parameters $\lambda^{ri}$ corresponding to the isometry transformations can also be written as:
\bea
\lambda^{ai} \rightarrow \lambda = \left(\begin{array}{cc}\epsilon^1 & \epsilon^2 \\
{\bar \epsilon}^1  & {\bar \epsilon}^2\end{array}\right) : \quad \overline{({ \lambda}^{ai})} =
\varepsilon_{a a^\prime}\varepsilon_{ii^\prime}\;{ \lambda}^{a^\prime i^\prime}~.
\eea
Then the corresponding isometry transformations become:
\bea
\delta z^i = \epsilon^i -\hat{\kappa}^2({\bar \epsilon}_jz^j)z^i +\hat{\kappa}^2(\epsilon^jz_j){\bar z}^i . \lb{S4tran}
\eea
The action \p{S4} describes a non-compact version of the 4-sphere; the genuine $S^4$ sigma model corresponds
to the change $\hat{\kappa}^2 \rightarrow -\hat{\kappa}^2$ in \p{S4} and \p{S4tran}. The same change
in eqs. \p{HP6} - \p{IdInv} (and $\hat{\kappa} \omega \rightarrow |\hat{\kappa}| \omega$ in eqs. \p{FhatF}, \p{hatFF}) gives rise to
$1D$ sigma model on the compact QK manifold $\mathbb{H}{\rm P}^n = Sp(1 +n)/[Sp(1) \times Sp(n)]\,$.

Before closing this Section, it is worth to give another form of the action \p{HP6}, in which
it involves the auxiliary fields $D(t)$ and $V^{(ij)}(t)$ \cite{Ivanov:1999vg}
\bea
S_{HP}' &=& \frac12 \int dt \Big\{ \dot{\hat{F}}^{ir} \dot{\hat{F}}_{ir} - \dot{f}^{ia} \dot{f}_{ia} + \frac12 D\left(f^{ia}f_{ia}
- {\hat{F}}^{ir} {\hat{F}}_{ir}
-\frac{2}{\hat{\kappa}^2}\right) \nn
&& - 2V^{ij}\left(\hat{F}^{r}_{(i} \dot{\hat{F}}_{j)r}  - {f}^{a}_{(i} \dot{f}_{j)r}\right)  - \frac{1}{\hat{\kappa}^2}\,V^{ij}V_{ij}\Big\}.\lb{Action-prime}
\eea
As it will become clear later, this action respects a local $Sp(1) \sim SU(2)$ invariance which can be used to fix
the $Sp(1)$ gauge $f^i_a = \sqrt{2}\, \delta^i_a\,\omega\,,$ where the factor $\sqrt{2}$ was inserted for further convenience.
Using this gauge, we can pass in \p{Action-prime} to the field $\omega(t)$, $f^{ia}f_{ia} = 4\omega^2,\;
\dot{f}^{ia} \dot{f}_{ia}  = 4\dot\omega \dot\omega\,.$
The $4D$ prototypes of the non-propagating fields $D$ and $V^{ik}$ come from the ${\cal N}=2, 4D$ Weyl multiplet \cite{MatN2}.
The field $D$ is just the Lagrange
multiplier producing the constraint to eliminate the field $\omega(t)$:
\be
\omega^2 - \frac14 {\hat{F}}^{ir} {\hat{F}}_{ir} -\frac{1}{2\hat{\kappa}^2} = 0 \; \Rightarrow \;\omega(t) =
{1\over \sqrt{2}\,|\hat{\kappa}|}\sqrt{1 + {\hat{\kappa}^2\over 2}\,
\hat{F}^2}\,,
\lb{omegaConstr}
\ee
while $V^{ik}$ is expressed by its algebraic equation of motion as
\be
V_{ij} = \hat{\kappa}^2\, \dot{{\hat{F}}}^{r}_{(i} {\hat{F}}_{j)r}\,.\lb{Current}
\ee
Substituting \p{omegaConstr} and \p{Current} back into \p{Action-prime} reproduces the $\mathbb{H}{\rm H}^n$ sigma model action \p{HP6}.
The $\mathbb{H}{\rm P}^n$ case is recovered by changing, in \p{Action-prime}, the sign before terms bilinear in $f^{ia}$ and $\dot{f}^{ia}$,
as well as by the replacement $\hat{\kappa}^2 \rightarrow - \hat{\kappa}^2\,$. Respectively, in \p{omegaConstr} and \p{Current}  one should replace $\omega^2 \rightarrow - \omega^2$,
and $\hat{\kappa}^2 \rightarrow - \hat{\kappa}^2\,$.

Finally, note that the action \p{Action-prime} (and its $\mathbb{H}{\rm P}^n$ counterpart) looks to be singular in the limit $\hat{\kappa} \rightarrow 0$
in contrast to \p{HP6} or \p{HP7}. However, this singularity
is fake. Redefining (in the $Sp(1)$ gauge),  $\omega$ as $\omega = \frac{1}{\sqrt{2}\hat{\kappa}} + \kappa\tilde{\omega}\,$,
we can represent the kinetic $\omega$
term in \p{Action-prime} as
\be
4\dot\omega\dot\omega  = 4\hat{\kappa}^2 \dot{\tilde{\omega}}\dot{\tilde{\omega}}\,,
\ee
while the coefficient of the field $D$ as
\be
4\omega^2 - {\hat{F}}^2 -\frac{2}{\hat{\kappa}^2} \; \Rightarrow  \;  4\sqrt{2}\tilde{\omega}
+ 4\hat{\kappa}^2\tilde{\omega}^2- \hat{F}\cdot \hat{F}.
\ee
Also, one redefines $V^{ik} = \hat\kappa \tilde{V}^{ik}$. After these redefinitions, the limit $\hat\kappa \rightarrow 0$
becomes well-defined.
In this limit, the fields $\tilde\omega$ and $\tilde{V}^{ik}$ fully decouple\footnote{The former algebraic constraint
becomes $\tilde\omega = \frac{1}{4\sqrt{2}}\hat{F}^2$, which is now just the definition of $\tilde\omega$.} and we are left with
the free kinetic term of $\hat{F}^{ir}$ as the only surviving one. An analogous reasoning applies to the $\mathbb{H}{\rm P}^n$ case.
Note that in the limit $\hat\kappa \rightarrow 0\,,$
\be
\sqrt{2}\,|{\hat \kappa}| \omega = 1\,. \label{limit}
\ee
This condition is also fulfilled in the general case of QK sigma models of refs. \cite{MatN2} - \cite{Ivanov:1999vg}.

\setcounter{equation}{0}

\section{QK supersymmetric mechanics: Geometric and group-theoretical setup}
\subsection{${\cal N} = 4\,, \,1D$ harmonic superspace and superfields}

We deal with one-dimensional ${\cal N}=4$ supersymmetry realized in $1D$ harmonic superspace.
We basically follow the notations and conventions of ref. \cite{ILe}.
The $1D$ harmonic superspace in the analytic basis amounts to the coordinate set\footnote{We denote
the complex time variable in the analytic basis
by the same letter as in the central basis, hoping that this will not cause a confusion. The relations to the central basis,
as well as the precise ${\cal N}=4$
transformation properties of the superspace coordinates, can be found in \cite{ILe} and, more recently, in \cite{DI1}.}
\bea
z := (t, \theta^+, \bar\theta^+, \theta^-, \bar\theta^-, w^\pm_i)\,, \quad w^+_iw^-_k - w^+_k w^-_i = \varepsilon_{ik}\,, \lb{1DHSS}
\eea
where $w^\pm_i$  are harmonics parametrizing the automorphism $SU(2)$ group. The analytic harmonic superspace is a subset of \p{1DHSS}
\bea
\zeta := (t, \theta^+, \bar\theta^+, w^\pm_i)\,. \lb{1DASS}
\eea
Both \p{1DHSS} and \p{1DASS} are closed under the appropriate realization of ${\cal N}=4$ supersymmetry.
In what follows, we will need the expressions for harmonic derivatives
\bea
&& D^{++} = \partial^{++} + 2i\theta^+\bar\theta^+ \partial_t + \theta^+\partial_{\theta^-} + \bar\theta^+\partial_{\bar\theta^-}\,, \lb{D++} \\
&& D^{--} =
\partial^{--} + 2i\theta^-\bar\theta^- \partial_t + \theta^-\partial_{\theta^+}
+ \bar\theta^-\partial_{\bar\theta^+}\,, \lb{D--} \\
&& [D^{++}, D^{--}] = D^0 = \partial^0 + \theta^+\partial_{\theta^+} + \bar\theta^+\partial_{\bar\theta^+}
- \theta^-\partial_{\theta^-} - \bar\theta^-\partial_{\bar\theta^-}\,, \lb{DDcom}
\eea
where
\be
\partial^{\pm\pm} = w^\pm_i \frac{\partial}{\partial w^\mp_i}\,, \quad \partial^0 =
w^+_i \frac{\partial}{\partial w^+_i} - w^-_i \frac{\partial}{\partial w^-_i}\,.
\ee
The covariant derivative $D^{++}$ preserves the analyticity: the result of its action on the analytic superfield $\Phi(\zeta)$,
i.e. $D^{++}\Phi(\zeta)$, is again
defined on the set $\zeta$, i.e. it is an analytic $1D$ superfield. It is not true for $D^{--}\Phi(\zeta)$ which is $1D$ harmonic superfield,
but not an analytic
superfield. The operator $D^0$ counts the external harmonic $U(1)$ charge of the general harmonic and harmonic-analytic  $1D$ superfields.
All superfields are assumed to
have a definite harmonic $U(1)$ charge and so are eigenfunctions of $D^0$.

The coordinate sets \p{1DHSS} and \p{1DASS} are closed and real with respect to the generalized $\,\widetilde{\,}\,$ - conjugation
\bea
\widetilde{t} = t\,, \quad \widetilde{\theta^{\pm}} = \bar\theta^{\pm}\,, \; \widetilde{\bar\theta^{\pm}} = - \theta^{\pm}\,, \quad
\widetilde{w^\pm_i} = w^{\pm\,i} = \varepsilon^{ik}{w^\pm_k}\,, \; \widetilde{w^{\pm\,i}} = - w^{\pm}_{i}\,.\lb{Tilde}
\eea

In what follows we will deal with two analytic superfields $q^{+ a}(\zeta)$ and $2n$ analytic superfields $\hat{Q}^{+ r}(\zeta)$,
with $a=1,2$ and $r = 1, \ldots 2n$ being,
respectively, the indices of the fundamental representations of some extra groups $Sp(1) \sim SU(2)$ and $Sp(n)$
commuting with supersymmetry.
These superfields are subjected to the tilde-reality conditions
\bea
\widetilde{q^+_a} = \varepsilon^{ab}q^+_b, \quad \widetilde{\hat{Q}^+_r} = \Omega^{rs}\,\hat{Q}^+_s\,, \lb{Reality}
\eea
where $\Omega^{rs}, \Omega^{rp}\Omega_{ps} = \delta^r_s\,,$ are skew-symmetric constant $Sp(n)$ invariant ``metrics''.
In the terminology of refs. \cite{MatN2,GBIO,GIOO,Ivanov:1999vg}
these superfields are $1D$ analogs of the compensating hypermultiplet and ``matter'' hypermultiplets, respectively.
In the case of simplest $\mathbb{H}{\rm H}^n$ and
$\mathbb{H}{\rm P}^n$ sigma models which
will be the main subject of consideration in the present paper, these superfields are subject to the simple harmonic constraints
\bea
D^{++}q^+_a(\zeta) = 0\,, \qquad  D^{++} \hat{Q}^+_r(\zeta) = 0\,. \lb{HSconstrLin}
\eea
The case of general QK sigma model (Sect. 5) corresponds to some nonlinear versions of these constraints.  The explicit solution of
\p{HSconstrLin} is as follows
\bea
&& q^{+\, a}(\zeta) = f^{ia}(t)w^+_i + \theta^+\chi^a(t) - \bar\theta^+\bar\chi^a(t) - 2i\theta^+\bar\theta^+ \dot{f}^{ia}(t)w^-_i\,,
\lb{qcomp} \\
&& \hat{Q}^{+\, r}(\zeta) = \hat{F}^{ir}(t)w^+_i + \theta^+\chi^r(t) - \bar\theta^+\bar\chi^r(t)
- 2i\theta^+\bar\theta^+ \dot{\hat{F}}^{ir}(t)w^-_i\,. \lb{Qcomp}
\eea
The superfield reality conditions \p{Reality} imply the following reality properties for the component fields:
\bea
\widetilde{f^+_a} = f^{+ a} \,\Leftrightarrow \, \overline{(f_{ia})} = f^{ia}\,, \; \overline{(f^{ia})} = f_{ia}\,; \qquad
\overline{(\chi_a)} = \bar{\chi}^a\,, \;\overline{(\chi^a)} = -\bar{\chi}_a
\eea
(and similar ones for $\hat{F}^{ir}, \chi^r$). It is assumed that the indices $a$ and $r$ are raised and lowered in the standard
way by the skew-symmetric tensors
$\varepsilon_{ab}, \varepsilon^{ab}$ and $\Omega_{rs}, \Omega^{rs}$. We observe that $q^{+ a}$ carries 4 real bosonic degrees of freedom
and $\hat{Q}^{+\, r}$ - $4n$ such degrees, total of $4(n+1)$ bosonic degrees.

One more relevant superfield will be the scalar real superfield $H(z)$ which collects the objects of ${\cal N}=4, 1D$ ``supergravity''.
It lives on the whole harmonic superspace \p{1DHSS}
and is subjected to the purely harmonic constraint
\be
D^{++}H = 0\,, \quad \widetilde{H} = H\,, \lb{Hconstr}
\ee
which actually means that $H$ in the central basis does not depend   on harmonics at all, and so
is the standard harmonic-independent ${\cal N}=4\,, 1D$ superfield.
In the analytic basis, the component structure of $H$ is as follows
\bea
H(z) &=& h + \theta^+\theta^- M - \bar\theta^+\bar\theta^- \bar M + \theta^+\bar\theta^- (\mu - i\dot{h})
+ \bar\theta^+\theta^- (\mu + i\dot{h}) \nn
&& + \,4i (\theta^+\bar\theta^+ w^-_iw^-_k - \theta^+\bar\theta^- w^-_iw^+_k - \theta^-\bar\theta^+ w^-_iw^+_k
+ \theta^-\bar\theta^- w^+_iw^+_k) L^{(ik)} \nn
&& +\,4\theta^+\bar\theta^+\theta^-\bar\theta^- [D + 2\dot{L}^{(ik)}w^+_iw^-_k]
\lb{BosH} \\
&& +\, (\theta^-w^+_i - \theta^+w^-_i)\phi^i - (\bar\theta^-w^+_i - \bar\theta^+w^-_i)\bar\phi^i
+ 4i \theta^-\bar\theta^- (\theta^+w^+_i \sigma^i
- \bar\theta^+w^+_i \bar\sigma^i) \nn
&& + \,2i\theta^+\bar\theta^+ [\theta^-w^-_i(2\sigma^i - \dot{\phi}^i)
-\bar\theta^-w^-_i(2\bar\sigma^i - \dot{\bar\phi}^i)]\,.\lb{FermH}
\eea
It includes eight bosonic fields $h(t), M(t), \bar M(t), \mu(t), D(t), L^{(ik)}(t)$ and eight fermionic fields
$\phi^i(t), \bar\phi^i(t), \sigma^i(t), \bar\sigma^i(t)$.
The conjugation rules for the bosonic fields are evident, while for the fermionic ones they read
\be
\overline{(\phi_i)} = \bar\phi^i\,, \quad \overline{(\sigma_i)} = \bar\sigma^i\,, \lb{RealHcomp}
\ee

\subsection{Minimal local ${\cal N}=4, 1D$ supersymmetry}
The ${\cal N}=2, 4D$ QK sigma models with the bosonic target dimension $4n$ arise when $n$ matter hypermultiplets
and one compensating hypermultiplet are
coupled to conformal ${\cal N}=2, 4D$ supergravity, i.e. to ${\cal N}=2, 4D$ Weyl multiplet \cite{MatN2}. The basic features
of the latter which are decisive for forming
the correct nonlinear QK sigma model Lagrangian is the presence of a scalar auxiliary field and a triplet of non-propagating gauge
fields associated with
a local $SU(2)$ symmetry as a part of conformal supergravity group. This latter $SU(2)$ symmetry can be used to fully gauge away
a triplet of scalar fields from
the set of four physical-dimension bosonic fields of the compensating hypermultiplet, while the remaining bosonic field
is eliminated by the constraint for which the scalar
auxiliary field serves as a Lagrangian multiplier. One more source of nonlinearity is the elimination of the $SU(2)$ gauge field
in terms of bosonic components
of the ``matter'' hypermultiplets by its algebraic equation of motion. In the context of  ${\cal N}=2, 4D$ HSS approach
all these steps are carefully explained
in \cite{Ivanov:1999vg}.

Taking this into account, as the first step in constructing QK ${\cal N}=4, 1D$ sigma model one should define the proper extension
of the rigid ${\cal N}=4, 1D$ supersymmetry, such
that it includes a local version of $SU(2)$ automorphism group. Once again, by analogy with the higher-dimensional cases,
it is natural to assume that this extension
preserves the harmonic ${\cal N}=4$ analyticity. Hence, the relevant subgroup of the general ${\cal N}=4, 1D$
superdiffeomorphism group should be given
by the following set of coordinate transformations
\bea
&& \delta t = \Lambda (\zeta)\,, \; \delta \theta^+ = \Lambda^+(\zeta)\,, \; \delta \bar\theta^+ = \bar\Lambda^+(\zeta)\,, \;
\delta w^+_i = \Lambda^{++}(\zeta) w^-_i\,,\;
\delta w^-_i = 0\,, \lb{AnalPres} \\
&&\delta \theta^- = \Lambda^-(z)\,, \quad  \delta \bar\theta^- = \bar\Lambda^-(z)\,, \lb{Neanal} \\
&& \widetilde{\Lambda} = \Lambda\,, \; \widetilde{\Lambda^\pm}  = \bar\Lambda^\pm\,, \;
\widetilde{\bar\Lambda^\pm}  = - \Lambda^\pm\,, \;\widetilde{\Lambda^{++}} = \Lambda^{++}\,. \lb{RealGroup}
\eea
For the time being, all these parameter superfunctions are arbitrary functions of their arguments. An asymmetric
form of the transformation
of the harmonic
variables in \p{AnalPres} has been postulated on the pattern of the higher-dimensional analogs of \p{AnalPres} and \p{Neanal}
(see, e.g., \cite{MatN2}) and their more direct ${\cal N}=8, 1D$ analog \cite{BISu}.

One way to go further is to covariantize the flat analyticity-preserving harmonic derivative $D^{++}$ by the appropriate
harmonic superfield  vielbeins,
$D^{++} \Rightarrow
\nabla^{++} = \partial^{++} + {\cal H}^{++}_{(t)}(\zeta)\partial_t +...$, so as to ensure the standard transformation
law for $\nabla^{++}$,
\bea
\delta \nabla^{++} = -\Lambda^{++} D^0\,.\lb{DTransf}
\eea
However, as distinct, e.g., from the analyticity-preserving ${\cal N}=8, 1D$ group considered in \cite{BISu}, in the case
under consideration the original transformation set
proves to be so powerful that it allows one to gauge all the newly introduced vielbein coefficients to their flat values,
$\nabla^{++} \Rightarrow D^{++}$, still leaving us
with a non-trivial residual infinite-dimensional superdiffeomorphism subgroup. To find its precise structure, it is simpler
to start just with the flat $D^{++}$, postulate
for it the transformation law \p{DTransf},
\bea
\delta D^{++} = -\Lambda^{++} D^0\,,\lb{DTransf1}
\eea
and derive the corresponding constraints on the parameter-functions introduced in \p{AnalPres} and \p{Neanal} from \p{DTransf1}.
These constraints are easily determined to be as follows
\bea
&& D^{++}\Lambda^{++} = 0\,, \lb{1} \\
&& D^{++}\Lambda -2i (\Lambda^+\bar\theta^+ + \theta^+\bar\Lambda^+) = 0\,, \lb{2} \\
&& D^{++}\Lambda^+ - \Lambda^{++}\theta^+ = 0\,, \quad D^{++}\bar\Lambda^+ - \Lambda^{++}\bar\theta^+ = 0\,, \lb{3} \\
&& D^{++}\Lambda^- - \Lambda^+ + \Lambda^{++}\theta^- = 0\,, \quad D^{++}\bar\Lambda^- - \bar\Lambda^+
+ \Lambda^{++}\bar\theta^- = 0\,. \lb{4}
\eea
Yet, the general solution of eqs. \p{1} - \p{3} is still rather broad, e.g. it involves {\it two} local $SU(2)$ groups,
one affecting harmonics and another combining
$\theta^+, \bar\theta^+$ into a doublet and not touching harmonic variables at all \footnote{It reflects
the property that the rigid ${\cal N}=4, 1D$ superalgebra possesses
$SU(2)\times SU(2)$ automorphisms which both can be realized in the superspaces \p{1DHSS} and \p{1DASS}.}.
To find the minimal solution, we require it to involve only
one local $SU(2)$ symmetry. The corresponding solution is uniquely given by the following expressions
\bea
\Lambda(\zeta) &=& 2b + 2i(\lambda^i w^-_i \bar\theta^+ - \bar\lambda^i w^-_i\theta^+)
+ 2i\theta^+\bar\theta^+ \tau^{(ik)}w^-_iw^-_k\,, \lb{a}\\
\Lambda^+(\zeta) &=& \lambda^i w^+_i +\theta^+[\dot{b} + \tau^{(ik)}w^+_i w^-_k]\,, \nonumber\\
\bar\Lambda^+(\zeta) &=& \bar\lambda^i w^+_i +\bar\theta^+[\dot{b} + \tau^{(ik)}w^+_i w^-_k]\,, \lb{b}\\
\Lambda^{++}(\zeta) &=& \tau^{(ik)}w^+_i w^+_k - 2i(\dot{\lambda}^i w^+_i \bar\theta^+ - \dot{\bar\lambda}^iw^+_i\theta^+)
- 2i\theta^+\bar\theta^+[\ddot{b} + \dot{\tau}^{(ik)}w^+_i w^-_k]\,, \lb{c}\\
\Lambda^-(\zeta) &=& \lambda^i w^-_i + \theta^+\tau^{(ik)}w^-_iw^-_k + \theta^-[\dot b - \tau^{(ik)}w^-_iw^+_k] \nn
&& -\,2i \theta^-(\bar\theta^+ \dot{\lambda}^iw^-_i- \theta^+ \dot{\bar\lambda}^iw^-_i) + 2i\theta^+\bar\theta^+ \theta^-\dot{\tau}^{(ik)}w^-_iw^-_k\,, \nn
\bar\Lambda^-(\zeta) &=& \bar\lambda^i w^-_i + \bar\theta^+\tau^{(ik)}w^-_iw^-_k + \bar\theta^-[\dot b - \tau^{(ik)}w^-_iw^+_k] \nn
&& +\,2i \bar\theta^-(\theta^+ \dot{\bar\lambda}^iw^-_i- \bar\theta^+ \dot{\lambda}^iw^-_i) + 2i\theta^+\bar\theta^+ \bar\theta^-
\dot{\tau}^{(ik)}w^-_iw^-_k\,. \lb{d}
\eea
Here, $b(t)$, $\tau^{(ik)}(t)$ and $\lambda^i(t),\;\; \bar\lambda^i(t)$ are arbitrary local parameters,
bosonic and fermionic, respectively. They satisfy the reality
conditions
\be
\bar b = b\,, \quad  \overline{(\tau^{ik})} = \tau_{ik}\,, \quad \overline{(\lambda_{i})} =\bar\lambda^i\,, \quad
\overline{(\lambda^{i})} = -\bar\lambda_i\,.
\ee
It is straightforward to check the closedness of the coordinate transformations \p{AnalPres}, \p{Neanal}
with the superparameters \p{a} - \p{d} with respect to the Lie brackets (transformations \p{AnalPres} with the parameters
\p{a} - \p{c} are closed on their own right,
indicating that the analytic harmonic superspace \p{1DASS} is closed under this local group). When
specializing to constant parameters, we come back to the semi-direct product of
rigid ${\cal N}=4, 1D$ supersymmetry and automorphism $SU(2)$ symmetry. The local  ${\cal N}=4, 1D$ supergroup
defined above is isomorphic to the
classical (having no central charges) ``small'' ${\cal N}=4$ superconformal symmetry\footnote{This realization of the ``small''
${\cal N}=4$ superconformal group in ${\cal N}=4, 1D$ harmonic superspace was found for the first time in \cite{DelSo}.
Being translated into the language
of standard ${\cal N}=4, 1D$ superspace, it coincides with that found in \cite{PaSo} in the context of spinning ${\cal N}=4$ particle.}.

{}For what follows, it will be important that $\Lambda^{++}(\zeta)$ can be represented as
\bea
\Lambda^{++} = D^{++}Z\,, \; Z = \tau^{(ik)}w^+_iw^-_k - \dot{b} + 2i (\bar\theta^+\dot{\lambda}^i - \theta^+\dot{\bar\lambda}^i)w^-_i
- 2i\theta^+\bar\theta^+\dot{\tau}^{(ik)}w^-_iw^-_k\,,\lb{Z}
\eea
with $\widetilde{Z} = Z\,$. Now it is easy to define the (passive) transformation properties of the superfields
$q^{+a}, Q^{+r}$ under the local ${\cal N}=4$ supersymmetry.
They are uniquely fixed by requiring the constraints \p{HSconstrLin} to be covariant:
\bea
\delta q^{+ a} = Z q^{+ a}\,, \quad \delta Q^{+ r} = Z Q^{+ r}\,. \lb{qQtransf}
\eea
Also, it is easy to establish the transformation properties of the harmonic derivative $D^{--}$:
\bea
\delta D^{--} = - (D^{--}\Lambda^{++})\, D^{--}\,, \quad D^{--}\Lambda^{++} = D^{++}\Lambda^{--}\,, \;
\Lambda^{--} := D^{--}Z\,.\lb{D--transf}
\eea
Note that \p{D--transf} and \p{1} imply the important consequence
\be
D^{--}\Lambda^{--} = 0. \lb{Lambda--constr}
\ee

Two more ingredients we will need are the integration measures over the full harmonic superspace \p{1DHSS}
and its analytic subspace \p{1DASS},
\be
\mu_H := dt  dw d^2\theta^+ d^2\theta^-\,, \quad \mu^{(-2)} := dt  dw d^2\theta^+\,.
\ee
They have the following simple transformation properties
\be
\delta \mu^{(-2)} = 0\,, \quad \delta\mu_H = \mu_H \,2Z\,.
\ee
\setcounter{equation}{0}
\section{Invariant actions}
Now we are ready to discuss the construction of the invariant actions. The simplest bilinear action of the superfields $q^{+ a}$
and $\hat{Q}^{+ r}$ invariant under the rigid ${\cal N}=4$ supersymmetry is (up to a renormalization factor)
\bea
S_{prob}  = \int \mu_H {\cal L}_{(2)}(q, \hat{Q})\,, \quad {\cal L}_{(2)}(q, \hat{Q}) = \Big(\gamma q^{+ a}q^{-}_a
- \hat{Q}^{+ r} \hat{Q}^{-}_{ r} \Big), \lb{Probe}
\eea
where
\bea
q^{-}_a := D^{--}q^+_a\,, \quad \hat{Q}^{-}_{ r} := D^{--}\hat{Q}^+_r\,, \lb{qminus}
\eea
and $\gamma = \pm 1$ (the role of the parameter $\gamma $ will become clear soon). Taking into account the transformation
properties \p{qQtransf}, \p{D--transf},
the variation of the action under the local ${\cal N}=4$ supergroup  is
\be
\delta S_{prob} = \int \mu_H (4 Z - D^{--}\Lambda^{++}){\cal L}_{(2)}(q, \hat{Q})\,,  \lb{deltaSprob}
\ee
where we also used that $q^{+a}q^+_a = Q^{+ r}Q^+_r = 0$ in virtue of antisymmetry of $\varepsilon_{ab}$ and $\Omega_{rs}$.
To construct the invariant action, we use the superfield $H$ defined in \p{Hconstr}, \p{BosH}, \p{FermH}   and ascribe to it the following (passive)
transformation law
\be
\delta H = (-4Z + 2D^{--}\Lambda^{++}) H.\lb{Htranpass}
\ee
The second weight term was added just to secure the covariance of the constraint in \p{Hconstr}.
Indeed, $D^{++}(-4Z + 2D^{--}\Lambda^{++}) = 0\,.$  Then the invariant action is given by the simple expression
\be
S = \int \mu_H H{\cal L}_{(2)}(q, \hat{Q})\,.\lb{LocalAct}
\ee
Its ${\cal N}=4$ variation is
\be
\delta S = \int \mu_H (D^{--}\Lambda^{++})H{\cal L}_{(2)}(q, \hat{Q}) = \int \mu_H (D^{++}\Lambda^{--})H{\cal L}_{(2)}(q, \hat{Q}),
\ee
and it vanishes after integration by parts with respect to $D^{++}$ and making use of the constraints \p{HSconstrLin}, \p{Hconstr}
together with the relations $D^{++}q^{-\, a} = q^{+\, a}$, $D^{++}\hat{Q}^{-\,r} = \hat{Q}^{+\, r}\,$.

One more relevant invariant (resembling the superfield Fayet-Iliopoulos terms) can be constructed from the single superfield $H$:
\be
S_{FI} \sim \int \mu_H\,\sqrt{H}\,. \lb{FI}
\ee
Indeed, $\delta \sqrt{H} =(-2Z + D^{--}\Lambda^{++})\sqrt{H}$ cancels the variation of the integration measure in \p{FI}, while the residual
contribution $\sim D^{--}\Lambda^{++} = D^{++}\Lambda^{--}$ vanishes after integrating by parts and taking account
of the constraint \p{Hconstr}.

The flat superfield action \p{Probe} enjoys linear realizations of $Sp(n,1)$ (for $\gamma=1$) and $Sp(n+1)$ (for $\gamma =-1$)
symmetries. Their $Sp(1)$ and $Sp(n)$ subgroups act as rotations of the superfields $q^{+a}$ and $\hat{Q}^{+ r}$
with respect to the indices $a$ and $r$, while
the coset  $Sp(n,1)/[Sp(n)\times Sp(1)]$ (or  $Sp(n+1)/[Sp(n)\times Sp(1)]$), respectively as
\bea
&&\delta q^{+ a} = -\lambda^{a s}\hat{Q}^+_s\,, \qquad \delta  \hat{Q}^{+ s} = \lambda^{as}q^+_a; \quad (\gamma = 1)\,, \lb{spn,1} \\
&&\delta q^{+ a} = \lambda^{a s}\hat{Q}^+_s\,, \qquad \delta  \hat{Q}^{+ s} = \lambda^{as}q^+_a; \quad (\gamma = -1)\,. \lb{spn+1}
\eea
The harmonic constraints \p{HSconstrLin} also respect these target space invariances.

These linear realizations become nonlinear in the locally ${\cal N}=4, 1D$ supersymmetric action \p{LocalAct} augmented
with the action \p{FI}, since the
${\cal N}=4, 1D$ ``supergravity''  superfield $H$ brings in the  auxiliary fields $D$ and $L_{ik}\,$. They, first, generate an
additional algebraic constraint relating the fields $f^{ +a}$ and $\hat{F}^{+ r}$ (under an assumption that $f^{ +a} = w^{+a} \,{\rm const} + \ldots$),
and, second, ensure some nonlinear addition
to the free action of $\hat{F}^{+ r}$. Besides, some local $1D$ symmetries can be gauge-fixed to further reduce the total dimension
of the target bosonic manifold
to that of the coset  $Sp(n,1)/[Sp(n)\times Sp(1)]$ or its compact analog $Sp(n+1)/[Sp(n)\times Sp(1)]$.

\subsection{Bosonic sector}

We will start from the superfield action
\bea
\mathbb{S}_{HP} = \frac18 \big(S + \beta\, S_{FI}\big) = \frac18\int \mu_H \Big[H\, {\cal L}_{(2)} + \beta \sqrt{H}\Big], \lb{startup}
\eea
where $\beta$ is a real parameter to be specified below and the numerical coefficient is chosen for further convenience.
Our first aim is to consider the bosonic sector of this action in order  to learn how it is related to $\mathbb{H}{\rm H}^n$ and $\mathbb{H}{\rm P}^n$ actions
discussed in Section 1.

The direct routine calculation yields
\bea
\mathbb{L}_{HP}^b &=& \frac12 h\,\Big(\dot{\hat{F}}^{ir} \dot{\hat{F}}_{ir} - \gamma\,\dot{f}^{ia} \dot{f}_{ia}\Big)
+ L_{ik}\Big(\hat{F}^{(i r} \dot{\hat{F}}^{k)}_{ r} - \gamma f^{(i a} \dot{f}^{k)}_{ a}\Big) \nn
&& +\, \frac1{4}\,D\Big(\gamma f^{ia}f_{ia} - \hat{F}^{i r}\hat{F}_{i r} +\frac{\beta}{\sqrt{h}}\Big) \nn
&& +\,\frac{\beta}{4}\,\frac{1}{\sqrt{h}h}\Big[L^{ik}L_{ik} - \frac1{8}\big(M\bar M + \mu^2 +\dot{h}^2\big)\Big].\lb{Lagrbos}
\eea

Inspecting  \p{Lagrbos}, we observe
\begin{itemize}
\item In order to have the correct normalization of the kinetic term of $\hat{F}^{ir}$ we are led to assume that the field $h$
starts with a positive constant,
\be
h(t) = h_0 + \Delta h(t)\,, \quad h_0 \geq 0\,. \lb{hconst}
\ee

\item The auxiliary fields $M, \bar M, \mu$ fully decouple.

\item Using the local $SU(2)$ gauge freedom with the parameter $\tau^{ik}(t)$ and making the standard (and very important) assumption that the
``superconformal compensator'' $q^{+ a}$ starts with a constant,
\be
q^{+ a} = \varepsilon^{ia} w^+_i\,{\rm const} + \ldots\,, \lb{ConfComp}
\ee
one can fully gauge away the symmetric part of $f^{ia}$,
so that
\be
f^i_a(t) = \sqrt{2}\delta^i_a \omega(t)\,.\lb{Sp1Gauge}
\ee

\item Using the local dilatation part of the $1D$ diffeomorphism parameter $b = b_0 + t \tilde{b}(t)$, we can  gauge $h(t)$ into its flat
value
\be
h(t) = h_0 =1\,.\lb{genDil}
\ee
\end{itemize}

Finally, exploiting these observations,
we see that the Lagrangian \p{Lagrbos} precisely coincides with that corresponding to \p{Action-prime} under the following identifications
\be
\gamma = 1\,, \quad \beta = -|\beta|, \;\;|\beta| = \frac{2}{\hat{\kappa}^2}\,, \quad L^{ik} = -V^{ik}\,. \lb{Identif1}
\ee
The choice $\gamma=-1$ corresponds to the maximally flat compact QK manifold $Sp(1+n)/$ $[Sp(1)\times Sp(n)]\,$. For the corresponding
algebraic constraint to be solvable in this case, one should choose
\be
\gamma=-1: \qquad  \beta = |\beta|\,, \quad  |\beta| = \frac{2}{\hat{\kappa}^2}\;\; \Longrightarrow \;\;
\omega ={1\over \sqrt{2}|\hat{\kappa}|}\sqrt{1 - {\hat{\kappa}^2\over 2}\,\hat{F}^2}\,.  \lb{Identif2}
\ee
Recall that our construction is by no means based on any ${\cal N}=2, 4D$ supergravity considerations, so $\hat\kappa$ is some free parameter
having no any conceivable connection with Einstein constant. The compact and non-compact cases are distinguished just by the sign of
``cosmological constant'' $\beta$.

Thus we have reproduced the action \p{Action-prime} of the nonlinear $\mathbb{H}{\rm H}^n$ sigma model, as well as its $\mathbb{H}{\rm P}^n$ counterpart,
starting from the superfield action \p{startup}
invariant under some minimal local ${\cal N}=4, 1D$ supersymmetry.  These actions are recovered in a particular
gauge with respect to the local time reparametrizations and local $SU(2)$ transformations. The central role in reproducing
these actions belongs to the auxiliary field $D(t)\,$, which is naturally provided by the $1D$ ``supergravity''
superfield $H$ and produces the necessary constraints.

Note that, as is seen from \p{Lagrbos}, the role of the field $h$ is similar to the einbein
in the models of relativistic (spinning) particles. Normally, after varying with respect to this field, the standard mass constraint
$\dot{f}\cdot \dot{f} - \dot{\hat{F}}\cdot \dot{\hat{F}} = 0$ is generated. In our case,
should we choose $\beta = 0$, something similar would happen, yielding
finally some variant
of relativistic  particle in the space with signature $(1, 4n)$ (since the triplet in $f^{ia}$ can still be gauged away by local $SU(2)$).
The total model (with all fermions included) then could be interpreted as a model of ${\cal N}=4$ spinning particle.
Perhaps it deserves a special attention\footnote{At this point, it is worth mentioning an analogy with ref. \cite{PaSo}
where a spinning particle coupled to a non-propagating ${\cal N}=4, 1D$ supergravity multiplet in ${\cal N}=4$ superspace was considered.
However, the ``matter'' there
was described by ${\cal N}=4$ multiplets (${\bf 1, 4, 3}$) as distinct from the multiplets (${\bf 4, 4, 0}$) in our case.
 Also, the authors of \cite{PaSo}
describe the ${\cal N}=4, 1D$ ``supergravity'' by a constrained (${\bf 1, 4, 3}$) super-einbein $E$  containing
no independent auxiliary field $D$ which plays
the crucial role in our construction. The ${\cal N}=4, 1D$ ``supergravity'' described by $H$ could be called ``non-minimal''
as opposed to
the ``minimal'' version exploited in \cite{PaSo}. Note that in \cite{GNPW,CLW}, ${\cal N}=4$ spinning particle models on QK manifolds (in the component approach)
were derived from the radial quantization of ${\cal N}=2, 4D$ supergravity BPS black holes. In these papers, a minimal set of ${\cal N}=4, 1D$ supergravity
fields generating the standard constraints was also used.}. However, in this paper we are interested in the case of $\beta \neq 0$. In this case,
the equation of motion for $h(t)$ {\it does not} generate any constraint: it just serves to eliminate the auxiliary field $D$.
In the gauge \p{genDil}, the $h$  equation reads
\bea
D = -\frac{3}{h_0}\,L^{ik}L_{ik} + \frac{(h_0)^{3/2}}{\beta} \Big(\dot{\hat{F}}^{ir}\dot{\hat{F}}_{ir}
- \gamma\dot{f}^{ia}\dot{f}_{ia}\Big). \lb{ExprD}
\eea
Since in the Lagrangian  \p{Lagrbos} $D$ appears multiplied by the algebraic constraint, on the shell of the latter it does not matter  whether or not $D$
can be solved for. Analogously, varying the total action with respect to the $1D$ gravitini fields $\phi^i, \bar\phi^j$ at $\beta\neq 0$ does not
generate any constraint (see the relevant remark in sect. 4.2).

It is also worth pointing out that the field $L^{ik}$ is just the $1D$ gauge field for the local $SU(2)$ symmetry, as follows from its $\tau^{ik}$ transformation (see eqs. \p{tauH}).
Once again, its equation of motion at $\beta \neq 0$ does not generate any constraint (as opposed to the spinning ${\cal N}=4$ particle case), but expresses $L^{ik}$
in terms of other fields.

As the last topic of this Section, let us derive the $Sp(1,n)/[Sp(1)\times Sp(n)]$ transformations (\ref{Ftransf}).
The superfield transformations \p{spn,1} imply the following transformations for the bosonic components of $q^+_a$ and $\hat{Q}^{+}_r$:
\bea
\delta f^{ia} = -\gamma|\hat{\kappa}|\lambda^{ra} \hat{F}^i_s\,, \quad \delta F^{ir} = |\hat{\kappa}|\lambda^{ra} f^i_a\,, \lb{TranIsom}
\eea
where, for further convenience, we rescaled the group parameters as $\lambda^{sa} \, \rightarrow \, |\hat{\kappa}|\lambda^{sa}$ and explicitly
introduced the parameter $\gamma = \pm 1$ in order to encompass both the non-compact and the compact cases.
Now we choose the $Sp(1)$ gauge \p{Sp1Gauge}. Requiring it to be preserved implies the modification of the transformations \p{TranIsom}
by adding the appropriate compensating local $SU(2)$ transformation
\bea
\tilde{\delta} f^{ia} = -\gamma |\hat{\kappa}|\lambda^{ra} \hat{F}^i_r + \tilde{\tau}^i_{\;j}f^{ja} \,, \quad
\tilde{\delta} F^{ir} = |\hat{\kappa}|\lambda^{ra} f^i_a  + \tilde{\tau}^i_{\;j}F^{jr}\,. \lb{TranIsom1}
\eea
The parameter $\tilde{\tau}^i_{\;j}$ is uniquely fixed from the requirement that $\tilde{\delta} f^{(ia)} =0$:
\bea
\tilde{\tau}^i_{\;j} = \gamma \frac{|\hat{\kappa}|}{2\sqrt{2}\, \omega} \big(\lambda^r_j \hat{F}^i_r + \lambda^{ri} \hat{F}_{jr} \big).
\eea
This implies
\bea
&&\tilde{\delta}\omega = \gamma \frac{|\hat{\kappa}|}{2\sqrt{2}}\,\lambda^{ri}\hat{F}_{ir}\,,\lb{tildeom} \\
&& \tilde{\delta} \hat{F}^{ir} = \sqrt{2}|\hat{\kappa}|\omega \lambda^{ri} + \gamma \frac{|\hat{\kappa}|}{2\sqrt{2}\,\omega}
\big(\lambda^{si} \hat{F}_{s j} + \lambda^{s}_{j} \hat{F}^i_{s} \big) \hat{F}^{jr}\,. \lb{hatFTransf}
\eea
For $F^{ir} = \frac{1}{\sqrt{2}|\hat{\kappa}|\omega}\hat{F}^{ir}$, these transformations provide
\be
\tilde{\delta} F^{ir} = \lambda^{ri} - \gamma\hat{\kappa}^2 \lambda_{sj}F^{is}F^{jr}\,,  \nonumber
\ee
that, for $\gamma =1$, coincides with (\ref{Ftransf}).

\subsection{Fermionic sector}
\noindent{\it A. Contribution from the 1st term in \p{startup}}.
\vspace{0.4cm}

The corresponding contribution to the off-shell Lagrangian is (for simplicity, we choose $\gamma = 1$)
\bea
\mathbb{L}^{f(1)}_{HP} &=& \frac{i}{4}h \left(\chi^a \dot{\bar{\chi}}_a - \dot\chi^a {\bar{\chi}}_a - \chi^r \dot{\bar{\chi}}_r
+ \dot\chi^r {\bar{\chi}}_r\right) \nn
&& +\, \frac{i}{2}\phi_i\left(\dot{f}^{ia}\bar{\chi}_a- \dot{\hat{F}}^{ir}\bar{\chi}_r\right) -
\frac{i}{2}\bar\phi_i\left(\dot{f}^{ia}{\chi}_a- \dot{\hat{F}}^{ir}{\chi}_r\right) \nn
&& +\,\frac{i}{2}\sigma_i\left({f}^{ia}{\bar\chi}_a- {\hat{F}}^{ir}{\bar\chi}_r\right) -
\frac{i}{2}\bar\sigma_i\left({f}^{ia}{\chi}_a- {\hat{F}}^{ir}{\chi}_r\right) \nn
&& +\, \frac{M}{8}\Big({\bar\chi}^a{\bar\chi}_a - {\bar\chi}^r{\bar\chi}_r\Big) - \frac{\bar M}{8}\Big({\chi}^a{\chi}_a - {\chi}^r{\chi}_r\Big)
+ \frac{\mu}{4}\Big({\bar\chi}^a{\chi}_a - {\bar\chi}^r{\chi}_r\Big). \lb{Ferm1}
\eea
\vspace{0.4cm}

\noindent{\it B. Contribution from 2nd term in \p{startup}}.
\vspace{0.4cm}

\bea
\mathbb{L}^{f(2)}_{HP} &=& \beta\, \frac{i}{32 h^{3/2}} \Big(\phi^i\dot{\bar\phi}_i - \bar\phi^i\dot\phi_i + 4\sigma ^i \bar\phi_i - 4\bar\sigma^i\phi_i\Big) \nn
&& +\, \beta\,\frac{3}{64 h^{5/2}}\Big(4iL^{ik}\phi_{(i}\bar\phi_{k)} + \frac{M}{2} \bar\phi^i\bar\phi_i - \frac{\bar M}{2} \phi^i\phi_i + \mu \phi^i\bar\phi_i \Big)\nn
&& +\, \beta\, \frac{15}{64\cdot 8}\,\frac{1}{h^{7/2}} (\phi^k\phi_k) (\bar\phi^i\bar\phi_i). \lb{Ferm2}
\eea

The full off-shell Lagrangian is a sum
\be
\mathbb{L}_{HP} = \mathbb{L}^{b}_{HP} + \mathbb{L}^{f(1)}_{HP} + \mathbb{L}^{f(2)}_{HP}\lb{BosFerm}
\ee

Now we observe that all fields contained in $1D$ ``supergravity'' superfield $H$ can be eliminated in one or another way.

\begin{itemize}

\item The field $h(t)$ can be gauged into a constant by using the gauge parameter $\tilde{b}(t)$.

\item The field $D(t)$ is the Lagrange multiplier for the constraint
\be
\gamma f^2 - \hat{F}^2 + \frac{\beta}{h^{1/2}} = 0\,.\lb{ConstrBase}
\ee

\item The auxiliary fields $L^{ik}, M, \bar M, \mu$ are eliminated by their algebraic equations of motion as

\bea
&& L^{ik} = -2 \frac{h^{3/2}}{\beta} \Big[\hat{F}^{(i r}\dot{\hat{F}}^{k)}_r - \gamma {f}^{(i a}\dot{f}^{k)}_a \Big] + \frac{3i}{8 h} \bar\phi^{(i}\phi^{k)}\,, \nn
&& M = \frac{4 h^{3/2}}{\beta} \Big(\chi^r\chi_r - \gamma \,\chi^a \chi_a\Big) -\frac{3}{4 h}\,\phi^i\phi_i\,, \nn
&& \bar M =\frac{4 h^{3/2}}{\beta} \Big(\gamma \,\bar\chi^a \bar\chi_a - \bar\chi^r\bar\chi_r\Big) +\frac{3}{4 h}\,\bar\phi^i\bar\phi_i\,,\nn
&& \mu =\frac{4 h^{3/2}}{\beta} \Big(\gamma \,\bar\chi^a \chi_a - \bar\chi^r \chi_r\Big) + \frac{3}{4 h}\, \phi^i \bar\phi_i\,. \lb{MbarMmu}
\eea

\item At last, the field $\sigma^i$ serves as a Lagrange multiplier giving rise to the constraint that expresses the $1D$ ``gravitino''
$\phi^i$ in terms of the ``matter fields'':

\bea
\phi^i = \frac{4 h^{3/2}}{\beta } \Big(\gamma f^{ia}\chi_a - \hat{F}^{i r}\chi_r \Big), \quad
\bar\phi^i = \frac{4 h^{3/2}}{\beta } \Big(\gamma f^{ia}\bar\chi_a - \hat{F}^{i r}\bar\chi_r \Big).\lb{phiConstr}
\eea
\end{itemize}

A good self-consistency check is to verify, e.g.,  that the $\lambda^i, \bar\lambda^i$ transformations of the left- and right-hand
sides of \p{phiConstr} computed according to \p{lambdaq}, \p{lambdaQ}, \p{lambdaH} precisely match each other with taking into account
the non-dynamical equations \p{ConstrBase} - \p{phiConstr}. This has been done with the positive result. Note that the equations of motion
for the ``gravitini'' $\phi^i, \bar\phi^j$ {\it do not} produce any constraints (as opposed to what happens in the spinning particle models), but merely
serve to eliminate the fermionic Lagrange multipliers $\sigma^i, \bar\sigma^i$ in terms of other fields. This is similar to the role of the equation
of motion for the einbein field $h(t)$ (recall the discussion around eq. \p{ExprD}). \\
\vspace{0.4cm}

\noindent{\it C. Gauges and peculiarities of realization of supersymmetry}.
\vspace{0.4cm}

The gauge already used above is $h = h_0 = const$. It fixes the dilatation parameter $b$ up to a constant shift of time.

The $Sp(1)$ gauge
\be
f^{(ia)} = 0 \qquad \Longleftrightarrow \qquad f^i_a = \sqrt{2} \delta^i_a \omega \lb{Sp1gauge}
\ee
implies that, for its preservation, the transformation of $f^{ia}$ under local supersymmetry \p{lambdaq} should be accompanied by the appropriate
induced $Sp(1)$ transformation
\be
\tilde{\delta}_\lambda f^{ia} = -\lambda^i\,\chi^a + \bar\lambda^i\,\bar\chi^a + \tau^i_{\;k (ind)}f^{ka}\,, \quad  \tau^{ik}_{(ind)} = \delta^k_a\,
\frac{1}{\sqrt{2}\omega} \big[\lambda^{(i}\chi^{a)} - \bar\lambda^{(i}\chi^{a)}\big].\lb{Inducedtau}
\ee
The same induced $Sp(1)$ transformation should accompany $\lambda$ transformations of any field carrying the doublet indices $i, j,...$, e.g., of  $\hat{F}^{ir}$.

One more gauge can be attained by properly fixing local supersymmetry \footnote{This gauge, together with $h= const$,
is known in $1D$ ``supergravity'' as the ``unitary gauge'' (see, e.g., \cite{VanHolt}). It radically simplifies everything.}
\bea
&& \phi^i = \bar\phi^i = 0 \quad \Rightarrow \label{GaugePhi} \\
&& \dot{\bar{\lambda}}^i = \frac{1}{4i h_0}\big(4iL^i_{\;k}\bar{\lambda}^k + M\,\lambda^i + \mu \,\bar{\lambda}^i\big) \lb{lambdaeq1} \\
&&\dot{\lambda}^i = \frac{1}{4i h_0}\big(4iL^i_{\;k}{\lambda}^k + \bar M\,{\bar\lambda}^i - \mu \,{\lambda}^i\big), \lb{lambdaeq2}
\eea
where it is assumed that the auxiliary fields $M, \bar M, \mu$ and $L^{ik}$ are given by their on-shell expressions \p{MbarMmu} in which $h = h_0$ and
the gauges \p{Sp1gauge} and  \p{GaugePhi} were also used. All things are drastically simplified in the gauge \p{GaugePhi}: the Lagrangian $\mathbb{L}^{f(2)}_{HP}$
is vanishing and the fermionic fields $\chi^a$ are eliminated in terms of $\chi^r$ in virtue of \p{phiConstr}
\be
\chi^a = -\frac{\gamma}{\sqrt{2}\omega} \delta^a_i\hat{F}^{ir}\chi_r\,, \quad \bar\chi^a = -\frac{\gamma}{\sqrt{2}\omega} \delta^a_i\hat{F}^{ir}\bar\chi_r\,.
\lb{omegatoomega}
\ee
Eqs. \p{lambdaeq1}, \p{lambdaeq2} imply that the residual supersymmetry transformations of the remaining quantities $\hat{F}^{ir}, \chi^r, \bar\chi^r$ are
given by the same laws \p{lambdaQ} (plus the additional $Sp(1)$ rotation of $\hat{F}^{ir}$ with the parameter $\tau^{ik}_{(ind)}$ defined in \p{Inducedtau}), in which
the local parameters $\lambda^i(t)$ and $\bar\lambda^i(t)$ are solved for from Eqs. \p{lambdaeq1}, \p{lambdaeq2} in terms of the remaining fields and constant
Grassmann parameters $\lambda^i_{(0)}$ and $\bar\lambda^i_{(0)}$. These expressions are not only nonlinear, but also non-local. Indeed, in the limit of vanishing fermions eq.
\p{lambdaeq2} becomes
\be
\dot{\lambda}^i = \frac{2 \sqrt{h_0}}{\beta}\,J^i_{\;k}{\lambda}^k\,, \quad   J^{ik} := -\hat{F}^{(i r}\dot{\hat{F}}^{k)}_r\,.
\ee
In the $2\times 2$ matrix notation it is solved through the ordered exponential
\be
{\lambda}^i(t) = [P\exp\{\frac{2 \sqrt{h_0}}{\beta}\,\int^t_{-\infty} d\tau J(\tau)\}]^i_{\;j}{\lambda}^j_{(0)}\,.\lb{Pexpon}
\ee

Perhaps, these non-localities can be avoided in the Hamiltonian approach, in which fields from the very beginning  are not subject to any gauges.

Finally, we present the full form of the residual fermionic on-shell Lagrangian $\mathbb{L}^f_{HP}$  written in terms  of the fields $\hat{F}^{ir}, \chi^r, \bar\chi^r$.
We use the expressions \p{omegatoomega} for the fermionic fields $\chi^a, \bar\chi^a$ and the above gauges. Then the total $\mathbb{L}^f_{HP}$ is given by the first and
last lines in \p{Ferm1}, as well as by the appropriate contribution from the last line in \p{Lagrbos}, where we should substitute \p{omegatoomega}
and the expressions for the auxiliary fields $M, \bar M$ and $\mu\,$. It reads
\bea
\mathbb{L}^f_{HP} &=& -\frac{i}{4}h_0\,G^{[sr]}\big(\dot{\chi}_s\bar{\chi}_r - \chi_s \dot{\bar{\chi}}_r\big)
+ \frac{i\gamma}{4 \omega^2}h_0 \hat{F}^{i(s}\dot{\hat{F}}^{r)}_i \chi_s \bar{\chi}_r \nn
&& -\,\frac{h_0^{3/2}}{2\beta}\big(G^{[sr]}G^{[fg]} + G^{[fs]}G^{[gr]}\big)\chi_s\chi_r \bar{\chi}_f\bar{\chi}_g\,,\lb{FermTot}
\eea
where
\be
G^{[sr]} := \Omega^{[sr]} + \frac{\gamma}{2\omega^2}\hat{F}^{i[s}{\hat{F}}^{r]}_i\,, \quad
\omega ={1\over \sqrt{2}|\hat{\kappa}|}\sqrt{1 + \gamma {\hat{\kappa}^2\over 2}\,\hat{F}^2}\,. \lb{G-def}
\ee

In order to reveal the geometric meaning of various terms in \p{FermTot} (with $\gamma = 1$ for simplicity), we firstly pass to another parametrization of our bosonic
manifold, by the fields $F^{ir}$
related to $\hat{F}^{ir}$ by the relations \p{FhatF}, \p{hatFF}. In this parametrization, we find that
\be
G^{[sr]} = g^{sr},
\ee
where $g^{sr}$ is defined in eq. \p{metr22} of Appendix C. Then, using the vielbein representation \p{def-e0}, \p{def-e} for $g^{rs}$,
redefining the fermionic fields as
\be
{\chi}_{\underline{p}} = e^s_{\underline{p}}\chi_s\,, \quad {\bar\chi}_{\underline{q}} = e^r_{\underline{q}}\bar\chi_r,
\ee
we can rewrite the terms bilinear in fermionic fields in \p{FermTot} (in the gauge $h =1$) as
\bea
-\frac{i}4\Big[\Omega^{\underline{q}\underline{u}}\big(\dot{\chi}_{\underline{q}}{\bar{\chi}}_{\underline{u}}
- {\chi}_{\underline{q}} \dot{{\bar{\chi}}}_{\underline{u}}\big) + \big(\Omega^{\underline{p}\underline{u}}\dot{e}^{\underline{q}}_s
e^s_{\underline{p}} + \Omega^{\underline{p}\underline{q}}\dot{e}^{\underline{u}}_s e^s_{\underline{p}}
- 2{\hat{\kappa}}^2 F^{i(s}\dot{F}_i^{r)} e^{\underline{q}}_s e^{\underline{u}}_r\big)
{\chi}_{\underline{q}}{\bar{\chi}}_{\underline{u}}\Big]. \lb{hatferm}
\eea
Now, substituting into \p{hatferm} the explicit expressions \p{def-e} for the vielbeins $e^{\underline{q}}_s, e^s_{\underline{p}}$,
after some algebra we can rewrite the Lagrangian \p{FermTot} (for $\gamma =1$) as
\bea
\mathbb{L}^f_{HP} = -\frac{i}{4}\,\Omega^{\underline{p}\underline{q}}\big(\nabla_{(t)}
{\chi}_{\underline{p}}{\bar{\chi}}_{\underline{q}} - {\chi}_{\underline{p}} \nabla_{(t)}{{\bar{\chi}}}_{\underline{q}}\big)
+\,\frac{1}{4} R^{({\underline{p}}{\underline{q}})\,({\underline{r}}{\underline{s}})}{\chi}_{\underline{p}}
{\chi}_{\underline{r}} {\bar{\chi}}_{\underline{q}}{\bar{\chi}}_{\underline{s}}\,,\lb{FermTot2}
\eea
where
\bea
\nabla_{(t)}{\chi}_{\underline{p}} = \dot{{\chi}}_{\underline{p}} - {\omega_{ir}}_{({\underline{p}}\;}^{\;\;\;\;\,{\underline{q}})}
\dot{F}^{ir}{{\chi}}_{\underline{q}}\,, \quad  R^{({\underline{p}}{\underline{q}})\,({\underline{r}}{\underline{s}})} = -\hat{\kappa}^2
\big(\Omega^{\underline{p}\underline{r}}\Omega^{\underline{q}\underline{s}}
+ \Omega^{\underline{p}\underline{s}}\Omega^{\underline{q}\underline{r}}\big).
\eea
The connection $\omega_{ir\; (\underline{p}\,\underline{q})}$ precisely coincides with the $Sp(n)$ part of the spin connection
on the considered  bosonic manifold (eq. \p{SpnCon}), while $R^{({\underline{p}}{\underline{q}})\,({\underline{r}}{\underline{s}})}$
is none other than the $Sp(n)$ part of the curvature tensor in the tangent-space representation (eq. \p{sp1spn2}).
The case with $\gamma = -1$ is recovered through
the substitution $\hat{\kappa}^2 \rightarrow - \hat{\kappa}^2$.

\section{Towards the generic ${\cal N}=4$ QK mechanics}
The generic HK ${\cal N}=4, 1D$ sigma model was obtained in \cite{DI1}  basically by passing from the linear harmonic constraints
of the type \p{HSconstrLin} to their nonlinear variant. The form of the latter coincides with the superfield equations of motion for
analytic hypermultiplet superfields in the most general HK ${\cal N}=2, 4D$ sigma model \cite{HSSb,HSShk1,HSShk2}. Yet a crucial difference between
the ${\cal N}=4, 1D$ and  ${\cal N}=2, 4D$ cases rooted in the fact that in the former case both linear and nonlinear harmonic constraints
do not imply the equations of motion for $4n + 4n$ physical bosonic and fermionic fields: these equations
follow from the invariant action which has the same universal form for both linear and nonlinear cases. The linear constraints
correspond to the free model, i.e. flat $4n$ dimensional target manifold; non-trivial HK models arise if and only  if the harmonic constraints
are nonlinear.

Here we will keep to the same strategy. Taking as an input the superfield equations of motion for the compensating
and physical hypermultiplet superfields following from the most general superfield action of QK sigma model in harmonic
superspace \cite{MatN2,GBIO,GIOO,Ivanov:1999vg}
we replace the flat harmonic constraints \p{HSconstrLin} by the following nonlinear ones
\bea
&& D^{++} q^{+ a} - \gamma \frac12 \frac{\partial}{\partial q^{+}_a}\Big[\hat{\kappa}^2 (w^-\cdot q^+)^2{\cal L}^{+4} \Big] = 0\,, \lb{qgen} \\
&& D^{++} \hat{Q}^{+ r} + \frac12 \frac{\partial}{\partial \hat{Q}^{+}_r}\Big[\hat{\kappa}^2 (w^-\cdot q^+)^2{\cal L}^{+4} \Big]
= 0\,,\lb{Qgen} \\
&& {\cal L}^{+4} \equiv {\cal L}^{+4} \Big(\frac{\hat{Q}^{ + r}}{\hat{\kappa} (w^-\cdot q^+)}, \frac{q^{ + a}}{(w^-\cdot q^+)}, w^-_i\Big),
\quad (w^-\cdot q^+) := w^-_{a}q^{+ a}\,. \lb{L+4}
\eea
The object ${\cal L}^{+4}$ is the renowned QK potential \cite{MatN2,GIOO}; the parameter $\hat\kappa$
is the contraction parameter to the general HK
case: when $\hat\kappa$ goes to zero, $\hat{\kappa} (w^-\cdot q^+) \rightarrow 1\,, \; \frac{q^{ + a}}{(w^-\cdot q^+)} \rightarrow w^{+a}$
and,
respectively, ${\cal L}^{+4} \rightarrow {\cal L}^{+4}(Q, w^+, w^-)$. In this limit \p{Qgen} becomes the nonlinear constraint describing
the most general HK ${\cal N}=4, 1D$ sigma model. The superfield $q^+_a$ fully decouples. It is important that the QK potential
does not involve any explicit $w^+$ harmonics.

The new constraints \p{qgen}, \p{Qgen} are covariant under local ${\cal N}=4, 1D$ supergroup. Indeed, the arguments of ${\cal L}^{+4}$
are manifestly invariant due to the transformation property
\be
\delta (w^-\cdot q^+) = Z\, (w^-\cdot q^+).
\ee
Then, taking into account that  $\delta (\frac{\partial}{\partial q^{+ a}}, \frac{\partial}{\partial \hat{Q}^{+ r}})
= -Z (\frac{\partial}{\partial q^{+ a}}, \frac{\partial}{\partial \hat{Q}^{+ r}})$, we see that the l.h.s.
of \p{qgen}, \p{Qgen} are homogeneously transformed as $q^{+ a}$ and $\hat{Q}^{+ r}$ themselves, like in the case
of linear constraints \p{HSconstrLin}.

It is surprising that  the invariant superfield action for the generic QK case is given by the same expression as for
the $\mathbb{H}{\rm H}^n$ and $\mathbb{H}{\rm P}^n$ cases
\bea
\mathbb{S}_{QK} = \frac18 \Big[\tilde{S} + \beta\, S_{FI}\Big] = \frac18\int \mu_H \Big[H\, \tilde{\cal L}_{(2)}
+ \beta \sqrt{H}\Big], \lb{startupG}
\eea
where $\tilde{\cal L}_{(2)}$ is formally given by the same bilinear expression as in \p{Probe}
\be
\tilde{\cal L}_{(2)} = \Big(\gamma q^{+ a}q^-_a - \hat{Q}^{+ r}\hat{Q}^-_r\Big)\,, \quad q^{- a} = D^{--} q^{+ a}\,,
\quad \hat{Q}^{- r} = D^{--}\hat{Q}^{+ r}\,.\lb{LagrG}
\ee
The crucial difference is that the analytic ${\cal N}=4, 1D$ superfields $q^{+ a}$ and $\hat{Q}^{+ r}$ are now
subject to nonlinear harmonic constraints \p{qgen}, \p{Qgen} and so possess more complicated $\theta^+, \bar\theta^+$ expansion.
With fermionic fields neglected, these expansions are
\bea
&& q^{+\, a}(\zeta) \, \Rightarrow \,  f^{+a}(t, w) + i\theta^+\bar\theta^+ A^{-a}(t, w)\,, \lb{qcompG} \\
&& \hat{Q}^{+\, r}(\zeta) \, \Rightarrow \, \hat{F}^{+ r}(t, w) + i\theta^+\bar\theta^+ \hat{A}^{- r}(t, w)\,. \lb{Qcomp1}
\eea
The functions $f^{+a}$, $\hat{F}^{+ r}$, $A^{-a}$ and $\hat{A}^{- r}$ are non-linearly expressed in terms of the ``central basis'' target coordinates
$f^{ia}(t), \hat{F}^{ir}(t)$ and their derivatives $\dot{f}^{ia}(t), \dot{\hat{F}}^{ir}(t)$ by non-dynamical harmonic equations which are the appropriate
$\theta, \bar\theta$ projections of the superfield constraints \p{qgen}, \p{Qgen}, like in the $\hat{\kappa} = 0$ limit \cite{DI1}. The equations for $f^{+ a},
\hat{F}^{+ r}$ are obtained from \p{qgen}, \p{Qgen} just by putting there $\theta = \bar\theta =0$, i.e. by the replacements $q^{+ a} \rightarrow f^{+ a},\;
\hat{Q}^{ + r} \rightarrow \hat{F}^{+ r}(t)\,, \; D^{++} \rightarrow \partial^{++}, \; {\cal L}^{+ 4} \rightarrow {\cal L}^{+ 4}|
:= {\cal L}^{+ 4}_{(\theta^+ = \bar\theta^+ = 0)}\,.$ The harmonic equations for $A^{- a}, \hat{A}^{- r}$ also directly follow from \p{qgen}, \p{Qgen}
\bea
&& \partial^{++} A^{- a} + 2\dot{f}^{+ a} - \gamma \frac12 \frac{\partial^2}{\partial f^{+}_b \partial f^+_a}\Big[\hat{\kappa}^2 (w^-\cdot f^+)^2{\cal L}^{+4}| \Big] A^-_b \nn
&&-\,\gamma \frac12 \frac{\partial^2}{\partial \hat{F}^{+}_r \partial f^+_a}\Big[\hat{\kappa}^2 (w^-\cdot f^+)^2{\cal L}^{+4}| \Big] \hat{A}^-_r   = 0\,, \lb{qgenA} \\
&& \partial^{++} \hat{A}^{- r} + 2\dot{\hat{F}}^{+ r} + \frac12 \frac{\partial^2}{\partial \hat{F}^{+}_s \partial \hat{F}^+_r}
\Big[\hat{\kappa}^2 (w^-\cdot f^+)^2{\cal L}^{+4}| \Big] \hat{A}^-_s \nn
&&+\,\frac12 \frac{\partial^2}{\partial {f}^{+}_b \partial \hat{F}^+_r}\Big[\hat{\kappa}^2 (w^-\cdot f^+)^2{\cal L}^{+4}| \Big] {A}^-_b   = 0\,.\lb{QgenA}
\eea
Note that the constraints \p{qgen}, \p{Qgen}, for any non-trivial ${\cal L}^{+ 4}$, explicitly break $Sp(n, 1)$ or $Sp(n+1)$ symmetry down to some its subgroup,
despite the fact that the action \p{startupG} still formally respects such a maximal symmetry.

The proof of local ${\cal N}=4, 1D$ invariance of the general action \p{startupG} is a bit tricky. The invariance of the second term in \p{startupG}
is checked in the same way as in the $\mathbb{H}{\rm H}^n$ ( $\mathbb{H}{\rm P}^n$) case. The ${\cal N}=4, 1D$ group variation of the first term can be reduced to the expression
\be
\delta \tilde{S} = - 2\int \mu_H \,\Lambda^{--} H \left(\gamma D^{++}q^{+ a}q^-_a - D^{++} \hat{Q}^{+ r} \hat{Q}^-_r \right),
\ee
where, in particular, the property $D^{--}\Lambda^{--}=0$ was used. Next, making use of the constraints \p{qgen}, \p{Qgen} and the evident property $D^{--}w^{-}_i = 0$,
we bring this variation to the form
\be
\delta \tilde{S} = - \int \mu_H \,\Lambda^{--} H \,D^{--}\left[\hat{\kappa}^2 (w^-\cdot q^+)^2 {\cal L}^{+4}\right].
\ee
This expression vanishes, as the integrand in it is a total harmonic derivative (because of the conditions $D^{--}H = D^{--}\Lambda^{--} = 0$).

As the last topic, we present the bosonic sector of the Lagrangian in the general QK action \p{startupG}
\bea
\mathbb{L}_{QK}^b &=& \frac12 h \int dw \Big(\gamma \dot{f}^{+ a}A^-_a - \dot{\hat{F}}^{+r} \hat{A}^-_r\Big)
+3\,L_{ik}\int dw\, w^{-i}w^{-k}\Big(\hat{F}^{+ r} \dot{\hat{F}}^{+}_{ r} - \gamma f^{+ a} \dot{f}^{+}_{ a}\Big) \nn
&& +\, \frac1{2}\,D \int dw \Big(\gamma f^{+a}\partial^{--}f_{a}^+ - \hat{F}^{+ r}\partial^{--}\hat{F}_{r}^+ + \frac{\beta}{2 \sqrt{h}}\Big) \nn
&& +\,\frac{\beta}{4h \sqrt{h}}\Big[L^{ik}L_{ik} - \frac1{8}\big(M\bar M + \mu^2 +\dot{h}^2\big)\Big].\lb{LagrbosQK}
\eea
Note that, while calculating the coefficient of $L_{ik}(t)$ in \p{LagrbosQK}, a non-trivial use of the constraints \p{qgen}, \p{Qgen},
\p{qgenA}, \p{QgenA} and the property
\be
\Big(f^{+ a}\frac{\partial }{\partial f^{+ a}} +  \hat{F}^{+ r}\frac{\partial}{\partial \hat{F}^{+ r}}\Big) {\cal L}^{+ 4}| = 0 \lb{ImpIden}
\ee
has been made. The basic steps of  deriving this troublesome term are described in Appendix B.
Note that the relation \p{ImpIden} just expresses the invariance of ${\cal L}^{+4}$ under the simultaneous constant rescalings
of the superfields $q^{+ a}$ and $\hat{Q}^{+ r}\,$. In the gauge $h= 1$, with identification $\beta = -\gamma \frac{2}{{\hat{\kappa}}^2}$ and after elimination of the auxiliary
fields as $M = \bar M = \mu =0$, the action \p{LagrbosQK} coincides with the direct $4D \rightarrow 1D$ dimensional
reduction of the most general bosonic QK sigma-model action derived in \cite{Ivanov:1999vg} from the superfield action of
${\cal N}=2, 4D$ supergravity coupled to the hypermultiplet matter \cite{MatN2}, \cite{GBIO}. This in fact proves that we have indeed constructed
here the most general set of QK ${\cal N}=4$ mechanics models parametrized by the QK potentials
${\cal L}^{+4}\Big(\frac{\hat{Q}^{ + r}}{\hat{\kappa} (w^-\cdot q^+)}, \frac{q^{ + a}}{(w^-\cdot q^+)}, w^-_i\Big)$.

As the final remark, let us note the existence of the locally ${\cal N}=4$ invariant super Wess-Zumino term given by the analytic
action
\bea
S^{(WZ)}_{QK} =
i\int \mu^{(-2)} {\cal L}^{+2}\Big(\frac{\hat{Q}^{ + r}}{\hat{\kappa} (w^-\cdot q^+)}, \frac{q^{ + a}}{(w^-\cdot q^+)}, w^-_i\Big). \lb{WZQK}
\eea
It is invariant due to the invariance of the analytic subspace integration measure. It is a direct generalization of analogous term in
the flat ${\cal N}=4, 1D$ supersymmetry \cite{DI1} and describes a coupling to an abelian background gauge field given on QK target
manifold.

\section{Summary and outlook}
In this paper, using the harmonic superspace methods, we proposed a new class of ${\cal N}=4$ supersymmetric mechanics models, with
the generic $1D$  QK sigma models as the bosonic core\footnote{To avoid a possible confusion, we should point out
that the supersymmetric mechanics models we have constructed are the standard ones, in the sense that the evolution parameter
is the worldline time $t$, associated with the canonical Hamiltonian. We do not know whether there is any link between our
construction and the Hamiltonian analogy
suggested in \cite{HamAnalog} and further worked out in \cite{Gun}. In the latter case an analog of the evolution parameter is a complex $\mathbb{CP}^1$
coordinate parametrizing the harmonic 2-sphere.}.  The basic distinguishing feature of these models is the {\it local}  ${\cal N}=4, 1D$
supersymmetry and {\it local} automorphism $SU(2)$ symmetry. For implementing these gauge symmetries it proved necessary to introduce
a generalized superfield vielbein $H$ and an extra ``compensating'' $({\bf 4, 4, 0})$  multiplet, besides $n$ such ``genuine'' multiplets
extending $4n$ bosonic target coordinates. These additional multiplets, after properly fixing local $1D$ supersymmetry,  ensure
the correct nonlinear actions with $4n$ dimensional QK manifolds as the bosonic targets. Our approach is intrinsically one-dimensional and does not require resorting
to any dimensional reduction procedure. We explicitly constructed the superfield and
component actions for the supersymmetric mechanics models based on $4n$ dimensional non-compact and compact homogeneous QK manifolds
$\mathbb{H}{\rm H}^n$ and  $\mathbb{H}{\rm P}^n$. The set of superfield constraints and the superfield action describing most general QK ${\cal N}=4$
mechanics were also presented. Like in the HK ${\cal N}=4$ mechanics models \cite{DI1}, the superfield action has the unique universal form
for any QK target manifold, the specificity of the given model being encoded in the non-linear harmonic constraints. The latter
can be chosen linear only for the maximal-dimension homogeneous QK manifolds just mentioned.

There remain many problems for further study. The most direct one is to construct Hamiltonian formulation of the models
presented, at least for the $\mathbb{H}{\rm H}^n$ and  $\mathbb{H}{\rm P}^n$ cases, including the construction of the relevant supercharges
and quantization. An interesting problem is to explicitly construct the $1D$ actions and supercharges for the full set of symmetric
QK manifolds, the Wolf spaces. The QK potentials ${\cal L}^{+ 4}$ for Wolf spaces were listed in \cite{GaOO,GIOO}.

One more proposal for further study is as follows. It is known that in the case of rigid ${\cal N}=4, 1D$ supersymmetry
the general action of $({\bf 4, 4, 0})$ multiplets
\cite{ILe,DI1,HKT},
\bea
S_{gen}(q) \sim \int \mu_H {\cal L}(q^{+ A}, q^{-B}, w^\pm), \quad A = 1, \ldots 2n\,,\lb{GenHGKT}
\eea
with $q^{+ A}$ being subjected to the linear harmonic constraints \p{HSconstrLin}, describes a particular class of ${\cal N}=4$
sigma models with the torsionful ``weak''
HKT geometry in the bosonic target space. It is interesting to find the
counterpart of this action with local ${\cal N}=4, 1D$ supersymmetry. Surprisingly, it is rather easy to achieve such a generalization.
Let us define
\bea
{X} := \sqrt{H}\,(q^{+ a}q^{-}_a)\,, \quad {Y} := \sqrt{H}\,(Q^{+ r}Q^-_r)\,, \quad  \delta_\Lambda ({X}, \;{Y}) = 0\,, \;
D^{\pm\pm}({X}, \;{Y}) = 0\,.
\eea
Then the action
\bea
S^{loc}(q, Q) = \int \mu_H \sqrt{H}{\cal F}({X}, {Y}, w^-) \lb{HKTloc}
\eea
is invariant under both local ${\cal N}=4, 1D$ supergroup and the internal symmetry group $Sp(1)\times Sp(n)\,$.
It provides a generalization of both
the actions \p{GenHGKT} and \p{startup}. Indeed in the rigid supersymmetry limit
$H=1$ the action \p{HKTloc} is reduced to a particular case of \p{GenHGKT} (with $Sp(1)\times Sp(n)$ isometry), while for the special choice
\bea
{\cal F}({X}, {Y}, w^-) = \gamma\,{X} - {Y} + \beta\,,
\eea
just to \p{startup} (modulo a numerical coefficient). It would be interesting to elaborate on the component form of \p{HKTloc}
and to identify the relevant target geometry. It is clear that the latter should be a type of ``QKT'' geometry.
A few explicit examples of such QKT geometries were earlier given in \cite{HOP}, proceeding from heterotic $({\bf 4, 0})$
supersymmetric $2D$ sigma models coupled to $2D$ supergravity.

One more generalization could be construction of  $1D$ sigma models invariant under an extended local ${\cal N}=4, 1D$ supersymmetry
including two local $SU(2)$ symmetries and related to ``large''  ${\cal N}=4,$ $1D$ superconformal group. A natural realization
of such an extended supersymmetry is achieved in the framework of bi-harmonic ${\cal N}=4, 1D$ superspace \cite{IvNied} which
thus provides an appropriate arena for such a generalization.

Altogether, it would be tempting to generalize to the QK case many other salient features
of the plethora of rigid ${\cal N}=4$ mechanics models (see, e.g., \cite{HKT}, \cite{ReviewSC} and refs. therein),
including a remarkable correspondence between such models and various
complexes in the differential geometries, superextensions of integrable Calogero-type models, etc. In particular,
it is of clear interest to construct locally supersymmetric versions of other off-shell ${\cal N}=4, 1D$ multiplets
(with the field contents $({\bf 3, 4, 1}), ({\bf 2, 4, 2}), ({\bf 1, 4, 3}), ({\bf 0, 4, 4})$) and to study the corresponding
${\cal N}=4$ mechanics models.

Finally, it is worth pointing out that  the problem of constructing QK ${\cal N}=8$ mechanics is still not solved. This task could be tackled either
through dimensional reduction from the locally ${\cal N}=2$ supersymmetric harmonic $4D$ hypermultiplet sigma models (as suggested in \cite{Gun})
or, following the line of the present paper, by independently defining the appropriate ${\cal N}=8, 1D$ supergravity and coupling to
it unconstrained analytic ${\cal N}=8, 1D$ hypermultiplets with an infinite number of auxiliary fields off shell.

\section*{Acknowledgements}
We thank Sergey Fedoruk for useful discussions. The work of EI was partly supported by RFBR grant No 15-02-06670 and a grant of the Ministry of Education and Science
of Russian Federation No 3.1386.2017. He is thankful to Murat G\"unaydin for
reviving his interest in supersymmetric QK mechanics. This study was started during the visit
of LM to the Bogoliubov Laboratory of Theoretical Physics. He is very indebted to  Bogoliubov LTP at JINR -Dubna, for hospitality and partial support.


\bigskip

\renewcommand\theequation{A.\arabic{equation}} \setcounter{equation}0
\subsection*{A\quad Local N = 4 transformations of the component fields. }
Here we present the explicit form of the local ${\cal N}=4$ transformations of the component fields in the $\theta$ expansions of
the basic superfields $q^{+ a}(\zeta), Q^{+ r}(\zeta)$ and $H(z)$ defined in \p{qcomp}, \p{Qcomp}, \p{BosH} and \p{FermH}.

The component transformations of some  ${\cal N}=4$ superfield $\Phi$ can be found out from the general form of its ``active'' transformation
\be
\delta^\star \Phi = W\,\Phi - \Lambda\partial_t\Phi - \Lambda^\pm\partial_{\theta^\pm}\Phi - \bar\Lambda^\pm\partial_{\bar\theta^\pm}\Phi
- \Lambda^{++}\partial^{--}\Phi\,,
\ee
where $W$ are the corresponding weight factors defined in \p{Z}, \p{qQtransf} and \p{Htranpass} and the superparameters $\Lambda\,, \;\Lambda^\pm\,, \;
\bar\Lambda^\pm$ and $\Lambda^{++}$ are given in \p{a} - \p{d}.

Using this general formula, we obtain the following transformations for the component fields:
\vspace{0.2cm}

\noindent{\it Superfields $q^{+ a}$ and $Q^{+ r}$:}
\vspace{0.2cm}

\bea
&& \delta_b f^{ia} = -2b\, \dot{f}^{ia} - \dot b\, f^{ia}\,, \; \delta_b\chi^a = -2b \dot{\chi}^a - 2\dot b\, \chi^a\,,  \;
\delta_b\,\bar{\chi}^a = -2b\, \dot{\bar\chi}^a - 2\dot b\, \bar{\chi}^a, \lb{qb} \\
&& \delta_\lambda f^{ia} = -\lambda^i\,\chi^a + \bar\lambda^i\,\bar{\chi}^a\,, \quad \delta_\lambda\chi^a = 2i\partial_t(\bar{\lambda}^i f_i^a)\,,
\;\delta_\lambda\bar\chi^a  = 2i\partial_t(\lambda^i f_i^a)\,, \lb{lambdaq}\\
&& \delta_\tau f^{ia} = \tau^i_{\;k}\,f^{ka}\,, \quad \delta_\tau \chi^a = \delta_\tau \bar{\chi}^a = 0\,. \lb{tauq}
\eea
The transformations of the fields $\hat{F}^{ir}, \chi^r$ have the same form:
\bea
&& \delta_b \hat{F}^{ir} = -2b \,\dot{\hat{F}}^{ir} - \dot b \,\hat{F}^{ir}\,, \; \delta_b\chi^r = -2b\, \dot{\chi}^r - 2\dot b\, \chi^r\,,  \;
\delta_b\bar\chi^r = -2b\, \dot{\bar\chi}^r - 2\dot b\, \bar{\chi}^r, \lb{Qb} \\
&& \delta_\lambda \hat{F}^{ir} = -\lambda^i\,\chi^r + \bar\lambda^i\,\bar{\chi}^r\,, \quad \delta_\lambda\chi^r = 2i\partial_t(\bar{\lambda}^i {\hat{F}}_i^r)\,,
\;\delta_\lambda\bar{\chi}^r  = 2i\partial_t(\lambda^i {\hat{F}}_i^r)\,, \lb{lambdaQ}\\
&& \delta_\tau  \hat{F}^{ir} = \tau^i_{\;k}\, \hat{F}^{kr}\,, \quad \delta_\tau \chi^r = \delta_\tau \bar{\chi}^r = 0\,. \lb{tauQ}
\eea

\vspace{0.1cm}

\noindent{\it Superfield $H$:}
\vspace{0.1cm}

\bea
&&\delta_b h = -2b\,\dot{h} + 4\dot{b}\,h\,, \quad \delta_b M = -2b\,\dot{M} + 2\dot{b}\,M\,, \quad
\delta_b \mu = -2b\,\dot{\mu} + 2\dot{b}\,\mu\,, \nn
&&\delta_b L^{(ik)} = -2b\,\dot{L}^{(ik)} + 2\dot{b}\,L^{(ik)}\,, \quad
\delta_b D = -2b\,\dot{D} + 2\partial_t(\ddot{b}h)\,, \nn
&& \delta_b \phi^i = -2b\,\dot{\phi}^i + 3\dot{b}\,\phi^i\,,
\quad \delta_b \sigma^i = -2b\,\dot{\sigma}^i + \dot{b}\,\sigma^i +  \ddot{b}\phi^i\,, \lb{bH}
\eea
\bea
&& \delta_\lambda h = \lambda^i \phi_i - \bar\lambda^i\bar\phi_i\,, \; \delta_\lambda  M = 2i\dot{\bar\lambda}^i \phi_i
+ i{\bar\lambda}^i (4\sigma_i - 2\dot{\phi}_i)\,, \; \delta_\lambda  \bar M = \overline{(\delta_\lambda  M)}\,, \nn
&& \delta_\lambda\mu = -i(\dot{\lambda}^i\phi_i +\dot{\bar\lambda}^i\bar\phi_i)
-{i}[\lambda^i(2\sigma_i - \dot{\phi}_i) + \bar\lambda^i(2\bar\sigma_i - \dot{\bar\phi}_i)] \nonumber \\
&& \delta_\lambda L^{(ik)} = \bar\lambda^{(i}\bar\sigma^{k)} - \lambda^{(i}\sigma^{k)}
-[\dot{\bar\lambda}^{(i}\bar\phi^{k)} - \dot{\lambda}^{(i}\phi^{k)}]\,, \nonumber \\
&& \delta_\lambda D = \lambda^i\dot{\sigma}_i - \bar\lambda^i\dot{\bar\sigma}_i
- \dot{\lambda}^i {\sigma}_i + \dot{\bar\lambda}^i{\bar\sigma}_i + \partial_t (\dot\lambda^i{\phi}_i
- \dot{\bar\lambda}^i{\bar\phi}_i)\,, \nn
&& \delta_\lambda \phi^i = \lambda^i M + \bar\lambda^i(\mu + i\dot{h}) + 4i \bar\lambda^k L^i_{\;k} - 4i\dot{\bar\lambda}^i h\,, \nn
&& \delta_\lambda \bar\phi^i = \bar\lambda^i \bar M - \lambda^i(\mu - i\dot{h}) + 4i \lambda^k L^i_{\;k} - 4i\dot{\lambda}^i h\,, \nn
&& \delta_\lambda \sigma^i = \dot{\lambda}^i M +
\dot{\bar\lambda}^i (\mu - i\dot{h}) + 2i \bar{\lambda}^k\dot{L}^i_{\;k} + i {\bar\lambda}^i D - 2i \ddot{\bar\lambda}^i h\,, \nn
&&\delta_\lambda {\bar\sigma}^i =  \dot{\bar\lambda}^i \bar M -
\dot{\lambda}^i (\mu + i\dot{h}) + 2i {\lambda}^k\dot{L}^i_{\;k} + i {\lambda}^i D - 2i \ddot{\lambda}^i h\,,
\lb{lambdaH}\\
&&{} \nn
&& \delta_\tau h = \delta_\tau M = \delta_\tau \mu = 0\,, \; \delta_\tau L^{(ik)} = h \dot{\tau}^{(ik)} - 2\tau^{(i m}L^{k)}_{\;m}\,, \;
\delta_\tau D = -2\dot{\tau}^{(ik)}L_{(ik)}\,, \nn
&& \delta_\tau \phi^i = \tau^i_{\;k}\phi^k\,, \quad \delta \sigma^i = \tau^i_{\;k}\sigma^k - \dot{\tau}^{(i k)}\phi_k\,. \lb{tauH}
\eea

The standard rigid ${\cal N}=4, 1D$ supersymmetry and $R$-symmetry $SU(2)$ transformations of the component fields
are recovered upon choosing the constant parameters in \p{qb} - \p{tauH}.

\renewcommand\theequation{B.\arabic{equation}} \setcounter{equation}0
\subsection*{B\quad Derivation of the second term in the general bosonic QK action (5.13). }

Here we briefly explain how to derive the second term in the general bosonic QK action \p{LagrbosQK}. While other terms are straightforwardly deduced from the
superfield action \p{startupG}, the derivation of the coefficient of $L_{ik}$ is rather tricky.

For simplicity we choose $\gamma = 1$, the general case can be recovered by multiplying all terms coming from $q^{+ a}$ by $\gamma$.
The parts of the bosonic Lagrangian  we are interested in come from the superfield expression
\be
\sim H\big(q^{+a}D^{--}q^+_a - \hat{Q}^{+ r}D^{--}\hat{Q}^{+}_{r} \big),
\ee
and are found to be
\bea
&&\mathbb{L}_{QK}^b(L^{ik}) = \frac18 \big( {\cal A} + {\cal B} + {\cal C} + {\cal D} +{\cal E} \big), \lb{LL} \\
&&{\cal A} := 2\dot{L}^{+-} \big(f^{+ a}\partial^{--}f^+_a - \hat{F}^{+ r}\partial^{--}\hat{F}^+_r\big), \quad
 {\cal B} := -{L}^{++} \big(f^{+ a}\partial^{--}A^-_a - \hat{F}^{+ r}\partial^{--}\hat{A}^-_r\big), \nn
&&{\cal C} := -2{L}^{--} \big(f^{+ a}\dot{f}^+_a - \hat{F}^{+ r}\dot{\hat{F}}^+_r\big), \quad
{\cal D} := - 2 {L}^{+=} \big(f^{+ a}A^-_a - \hat{F}^{+ r}\hat{A}^-_r\big), \nn
&&{\cal E} := -{L}^{++} \big(A^{- a}\partial^{--}f^+_a - \hat{A}^{- r}\partial^{--}\hat{F}^+_r\big). \lb{ABCDE}
\eea
We omit the integral over harmonics in  \p{LL}, but it is implicitly assumed, so we can integrate by parts with respect to harmonic derivatives.

The term  ${\cal C}$ already has the needed structure, so we apply to other terms. It will be shown later that the term ${\cal A}$ is a total harmonic derivative.
The sum of terms ${\cal B}$, ${\cal D}$ and ${\cal E}$, by representing $L^{++} = \partial^{++} L^{+-}$ and integrating by parts with respect to the harmonic derivatives,
can be reduced to the expression
\bea
{\cal B} + {\cal D} +{\cal E} \,&\Rightarrow &  \, 2 L^{+ -} \big[A^{-a}\partial^{--}(\partial^{++} f^+_a) -
\hat{A}^{- r}\partial^{--}(\partial^{++} \hat{F}^+_r) \nn
&& -\, \partial^{--} f^{+ a}\partial^{++} A^-_a  +  \partial^{--} \hat{F}^{+ r}\partial^{++} \hat{A}^-_r - f^{+ a} A^-_a
+ \hat{F}^{+ r} \hat{A}^-_r \big]. \lb{BDE}
\eea
Next, we eliminate $\partial^{++} f^+_a, \partial^{++} \hat{F}^+_r$ and $\partial^{++} A^-_a, \partial^{++} \hat{A}^-_r$
from the constraints \p{qgen}, \p{Qgen} and \p{qgenA}, \p{QgenA},
after which \p{BDE} is simplified to the form
\bea
{\cal B} + {\cal D} +{\cal E} \, \Rightarrow   \, 2 L^{+ -} \big[ 2\partial^{--} f^{+ a}\dot{f}^+_a
- 2 \partial^{--} \hat{F}^{+ r}\dot{\hat{F}}^{+}_{r}
- f^{+ a} A^-_a + \hat{F}^{+ r} \hat{A}^-_r \big].\lb{BDE2}
\eea
Then, using the identity
$$
 2\partial^{--} f^{+ a}\dot{f}^+_a = \partial^{--} ( f^{+ a}\dot{f}^+_a) + \partial_t ( \partial^{--} f^{+ a}{f}^+_a)
$$
together with an analogous one for $\hat{F}^{+ r}$, integrating by parts with respect to the harmonic derivatives and $\partial_t$,
we transform \p{BDE2} to the sum
\bea
2\big[{L}^{--} \big(f^{+}_{ a}\dot{f}^{+a} - \hat{F}^{+}_{ r}\dot{\hat{F}}^{+r}\big)
+ {\dot{L}}^{+-} \big(f^{+ a}\partial^{--}{f}^+_a - \hat{F}^{+ r}\partial^{--}{\hat{F}}^+_r\big)
- L^{+-}\big(f^{+ a} A^-_a  - \hat{F}^{+ r} \hat{A}^-_r \big)\big]. \nonumber
\eea
The first term in this sum is ``good'', while the second one, through integrating by parts, can be reduced to the expression
$$
{\dot{L}}^{--} \big(\partial^{--}f^{+ a}\partial^{++} f^+_a - \partial^{--}\hat{F}^{+ r}\partial^{++} \hat{F}^+_r \big).
$$
After using the constraints \p{qgen}, \p{Qgen},  the expression within the brackets proves to be proportional to $\partial^{--} X^{+ 4}$,
with
$X^{+ 4} := \hat{\kappa}^2(w^-\cdot f^+)^2 {\cal L}^{+4}$, so this expression is a total harmonic derivative because
of the property $\partial^{--} L^{--} = 0\,$.
The same is true of course for the term ${\cal A}$ in \p{ABCDE}.

It remains to inspect the third term in the above sum. Integrating by parts, it is reduced to
\bea
L^{--} \big(\partial^{++} f^{ + a} A^-_a - \partial^{++} \hat{F}^{ + r} \hat{A}^-_r + f^{ + a}\partial^{++} A^-_a
- \hat{F}^{ + r} \partial^{++}\hat{A}^-_r\big). \lb{Third}
\eea

We employ the constraints \p{qgen}, \p{Qgen} and \p{qgenA}, \p{QgenA} once again and, finally, make use of the relations
\bea
\Big(1 - f^{+ a}\frac{\partial}{\partial f^{+ a}} - \hat{F}^{+ r}\frac{\partial}{\partial \hat{F}^{+ r}}  \Big)\frac{\partial X^{+ 4}}{\partial f^{+}_b} =
\Big(1 - f^{+ a}\frac{\partial}{\partial f^{+ a}} - \hat{F}^{+ r}\frac{\partial}{\partial \hat{F}^{+ r}}  \Big)
\frac{\partial X^{+ 4}}{\partial \hat{F}^{+}_s} = 0\,,
\eea
which are consequences of the relation \p{ImpIden}. As the result, \p{Third} becomes
\bea
2 {L}^{--} \big(f^{+}_{ a}\dot{f}^{+a} - \hat{F}^{+}_{ r}\dot{\hat{F}}^{+r}\big)
\eea
and we finally obtain that, up to a total derivative,
\bea
\mathbb{L}_{QK}^b(L^{ik}) = \frac18 \big({\cal A} +{\cal B} +{\cal C} + {\cal D} +{\cal E}\big) =
\frac34\,{L}^{--} \big(f^{+}_{ a}\dot{f}^{+a} - \hat{F}^{+}_{ r}\dot{\hat{F}}^{+r}\big),
\eea
that yields just the second term in \p{LagrbosQK}.

\renewcommand\theequation{C.\arabic{equation}} \setcounter{equation}0
\subsection*{C\quad Geometry of the homogeneous QK manifolds  $Sp(1,n)/[Sp(1)\times Sp(n)]$
and  $Sp(1+n)/[Sp(1)\times Sp(n)]$. }
In this Appendix we present the basic geometric quantities for the homogeneous QK manifold $Sp(1,n)/[Sp(1)\times Sp(n)]$
and its compact analog $Sp(1+n)/[Sp(1)\times Sp(n)]$.

For our purposes, it will be more convenient to employ the parametrization \p{HP7}, in which the target metric and its inverse read
\bea
g_{ir\,js} = \varepsilon_{ij}\,a\, g_{rs}\,, \qquad g^{ir\,js} = \varepsilon^{ij}\,a^{-1}\, g^{rs}\,, \lb{metr2}
\eea
with
\bea
a = \frac{1}{1 - \frac{{\hat\kappa}^2}{2} {F}^2}\,, \; g_{rs} = \Omega_{rs} + {{\hat\kappa}^2}\,a \,F^t_r F_{t s}\,,\;g^{rs}
= \Omega^{rs} + {{\hat\kappa}^2}\,F^{t r} F_{t}^{s}\,, \;
g^{rs}g_{su} = \delta^r_u. \lb{metr22}
\eea
The corresponding vielbeins are defined by
\bea
&& g_{ir\,js} = e_{ir}^{\;k\underline{p}}\,\varepsilon_{kl}\Omega_{\underline{p}\underline{q}}\,e_{js}^{\;l\underline{q}}\,, \qquad
g^{ir\,js} = e^{ir}_{\;k\underline{p}}\,\varepsilon^{kl}\Omega^{\underline{p}\underline{q}}\,e^{js}_{\;l\underline{q}}\,, \nn
&& e_{ir}^{\;k\underline{p}} =\delta^k_i\, a^{1/2}\, e^{\underline{p}}_r\,, \qquad   e^{ir}_{\;k\underline{p}} =
\delta^i_k\, a^{-1/2}\, e^r_{\underline{p}}\,, \lb{def-e0} \\
&& e^{\underline{p}}_r = \delta^{\underline{p}}_r + {{\hat\kappa}^2}\frac{a}{1 + a^{1/2}}\,F^{t\underline{p}} F_{tr}\,, \qquad
e_{\underline{p}}^r = \delta_{\underline{p}}^r  - {{\hat\kappa}^2}\frac{a^{1/2}}{1 + a^{1/2}}\,F^{tr} F_{t\underline{p}}\,,\nn
&& g_{rs} = e^{\underline{p}}_r\,\Omega_{\underline{p}\underline{q}}\,e^{\underline{q}}_s\,,\;\;
g^{rs} = e^r_{\underline{p}}\,\Omega^{\underline{p}\underline{q}}\,e_{\underline{q}}^s\,, \qquad  e^s_{\underline{p}}\,e^{\underline{p}}_r
= \delta^s_r\,, \; e^{\underline{p}}_r\, e^r_{\underline{q}} = \delta^{\underline{p}}_{\underline{q}}\,. \lb{def-e}
\eea

The Levi-Civita connection for the metric \p{metr2},
$$
\Gamma^{lr}_{\;\;kq\,js} = \frac12\,g^{lr\,l'r'}\big(\partial_{kq} g_{l'r'\,js} +\partial_{js} g_{l'r'\, kq} - \partial_{l'r'} g_{kq\,js}\big),
$$
has the simple form
\bea
\Gamma^{lr}_{\;\;kq\,js} = {\hat\kappa}^2 a \big(\delta^l_j \delta^r_q\,F_{ls} + \delta^l_k \delta^r_s\,F_{jq}\big). \lb{L-C}
\eea
The Riemann curvature tensor,
$$
R^{lr}_{\;\;js \,iu\, kq} = \partial_{iu}\Gamma^{lr}_{\;\;kq\,js} - \partial_{kq}\Gamma^{lr}_{\;\;iu\,js} + \Gamma^{lr}_{\;\;iu \,l'r'} \Gamma^{l'r'}_{\;\;kq \,js} -
\Gamma^{lr}_{\;\;kq \,l'r'} \Gamma^{l'r'}_{\;\;iu \,js}\,,
$$
is also easy to calculate:
\bea
R^{lr}_{\;\;js \,iu\, kq} = {\hat\kappa}^2 \big(\delta^l_j \delta^r_q\, g_{ks \, iu} + \delta^l_k \delta^r_s\, g_{jq \, iu}
-  \delta^l_j \delta^r_u\, g_{is \, kq}
- \delta^l_i \delta^r_s\, g_{ju \, rq}\big), \lb{RT1}
\eea
whence
\bea
R_{lr\,js \,iu\, kq} = -{\hat\kappa}^2 a^2 \big[ \varepsilon_{lj} \varepsilon_{ki}\,
( g_{rq}g_{su} + g_{ru}g_{sq} ) + g_{rs}g_{uq}(\varepsilon_{li} \varepsilon_{jk} +
\varepsilon_{lk} \varepsilon_{ji}) \big]. \lb{RT2}
\eea
Using the relations \p{def-e}, it is easy to lift \p{RT2} to the tangent space
\bea
R^{(\rm tg)}_{l\underline{r}\,j\underline{s} \,i\underline{u}\, k\underline{q}} =
\varepsilon_{lj}\varepsilon_{ik}\, R_{(\underline{r}\underline{s})\,(\underline{u}\underline{q})} +
\Omega_{\underline{r}\underline{s}}\Omega_{\underline{u}\underline{q}}\,R_{(lj)\,(ik)}\,,\lb{sp1spn}
\eea
with
\bea
R_{(\underline{r}\underline{s})\,(\underline{u}\underline{q})}
=-{\hat\kappa}^2\big(\Omega_{\underline{r}\underline{u}}\Omega_{\underline{s}\underline{q}}
+ \Omega_{\underline{r}\underline{q}}\Omega_{\underline{s}\underline{u}}\big), \quad R_{(lj)\,(ik)} =
-{\hat\kappa}^2\big(\varepsilon_{li}\varepsilon_{jk} +\varepsilon_{lk}\varepsilon_{ji}\big).\lb{sp1spn2}
\eea
For completeness, we also present Ricci tensor and the scalar curvature
\bea
R^{(\rm tg)}_{(j\underline{s}\;k\underline{q})}  = -  2(2+n){\hat\kappa}^2\,\varepsilon_{jk}\Omega_{\underline{s}\underline{q}}\,, \quad
R = -8 n(2 + n){\hat\kappa}^2\,. \lb{scal}
\eea

The last topic will be the expression for the spin connection. It is calculated by the general formula
\bea
\omega_{ir\;[j\underline{p}\;k\underline{s}]} = e_{\,lq \, j\underline{p}} \Big(\partial_{ir} e^{lq}_{\;k\underline{s}}
+ \Gamma^{lq}_{\;\;ir\, l'r'}e^{l'r'}_{\;k\underline{s}}\Big),
\eea
where vielbeins are defined in \p{def-e0}, \p{def-e}. After rather tedious calculations,
we obtain the following splitting of the spin connection into its $Sp(1)$ and $Sp(n)$ components
\bea
\omega_{ir\;[j\underline{p}\;k\underline{s}]}  = \Omega_{\underline{p}\underline{s}}\, \omega_{ir\,(jk)}
+ \varepsilon_{jk}\omega_{ir\,(\underline{p}\,\underline{s})}\,, \lb{SCon}
\eea
where
\bea
\omega_{ir\,(jk)} &=& -\frac{{\hat\kappa}^2}{2} a \big( \varepsilon_{ij} F_{rk} + \varepsilon_{ik} F_{rj}\big), \lb{Sp1Con} \\
\omega_{ir\,(\underline{p}\,\underline{s})} &=& {\hat{\kappa}}^2 \,\frac{a}{1 + a^{1/2}} \Big\{\Omega_{\underline{p}{r}} F_{i\underline{s}}
+\Omega_{\underline{s}{r}} F_{i\underline{p}} \nn
&&+\, \frac{{\hat{\kappa}}^2}{2}\frac{a}{1 + a^{1/2}}
\Big[F_{i\underline{p}}(F^l_{\underline{s}}F_{lr}) + F_{i\underline{s}}(F^l_{\underline{p}}F_{lr})\Big]\Big\}.
\lb{SpnCon}
\eea

The case of the compact QK manifold $\mathbb{H}{\rm P}^n = Sp(1+n)/[Sp(1)\times Sp(n)]$ is recovered by the substitution
${\hat{\kappa}}^2 \rightarrow - {\hat{\kappa}}^2$
in the formulas above.
\providecommand{\href}[2]{#2}\begingroup\raggedright\endgroup


\begin{thebibliography}{99}

\bibitem{Witt}
E.\,Witten, {\it Dynamical breaking of supersymmetry},
Nucl. Phys. B {\bf 188} (1981) 513-554, doi:10.1016/0550-3213(81)90006-7.

\bibitem{cole}
R.A.\,Coles, G.\,Papadopoulos, {\it The Geometry of the one-dimensional supersymmetric nonlinear sigma models},
Class. Quant. Grav. {\bf 7} (1990) 427-438, doi:10.1088/0264-9381/7/3/016.

\bibitem{GPS}
G.W.\,Gibbons, G.\,Papadopoulos, K.S.\,Stelle, {\it HKT and OKT geometries on soliton black hole moduli spaces}, Nucl. Phys. B {\bf 508} (1997) 623-658,
doi:10.1016/S0550-3213(97)00599-3, {\tt arXiv:hep-th/9706207}.

\bibitem{Hull}
C.M.\,Hull,
{\it The Geometry of supersymmetric quantum mechanics}, {\tt arXiv:hep-th/9910028}.

\bibitem{HK1}
L.\,Alvarez-Gaum\'e, D.Z.\,Freedman, {\it Ricci-flat K\"ahler manifolds and supersymmetry}, Phys. Lett. B {\bf 94} (1980) 171-173, doi:10.1016/0370-2693(80)90850-3.

\bibitem{HK2}
L.\,Alvarez-Gaum\'e, D.Z.\,Freedman, {\it Geometric structure and ultraviolet finiteness in the supersymmetric $\sigma$ model}, Commun. Math. Phys. {\bf 80} (1981)
443-451, doi:10.1016/0003-4916(81)90006-3.

\bibitem{HSSa}
A. Galperin, E. Ivanov, S. Kalitzin, V. Ogievetsky and E. Sokatchev, {\it Unconstrained N=2
matter, Yang-Mills and supergravity theories in harmonic superspace}, Class. Quant. Grav.
{\bf 1} (1984) 469-498 [Corrigendum ibid. 2 (1985) 127], doi:10.1088/0264-9381/1/5/004.

\bibitem{HSSb}
A.S.\,Galperin, E.A.\, Ivanov, V.I.\, Ogievetsky, and E.S.\, Sokatchev, {\it Harmonic superspace},
Cambridge, UK: Univ. Pr. (2001) 306 p, doi:10.1017/CBO9780511535109.

\bibitem{HSShk1}
A.\,Galperin, E.\,Ivanov, V.\,Ogievetsky, E.\,Sokatchev, {\it Hyper-K\"ahler metrics and harmonic superspace}, Commun. Math. Phys. {\bf 103} (1986) 515-526,
doi:10.1007/BF01211764.

\bibitem{GIOT}
A.\,Galperin, E.\,Ivanov, V.\,Ogievetsky, P.K.\,Townsend, {\it Eguchi-Hanson type metrics from harmonic superspace}, Class. Quant. Grav. {\bf 3} (1986) 625-633,
doi:10.1088/0264-9381/3/4/017.

\bibitem{HSShk2}
A.\,Galperin, E.\,Ivanov, V.\,Ogievetsky, E.\,Sokatchev, {\it Gauge field geometry from complex and
harmonic analyticities. II. Hyper-K\"ahler case}, Ann. Phys. {\bf 185} (1988) 22-45, doi:10.1016/0003-4916(88)90257-6.

\bibitem{HKsqm}
A.\,Kirchberg, J.D.\,Lange, A.\,Wipf, {\it Extended supersymmetries and the Dirac operator}, Ann. Phys. {\bf 315} (2005) 467-487,
doi:10.1016/j.aop.2004.08.006, {\tt arXiv:hep-th/0401134}.

\bibitem{DI1}
F.\,Delduc, E.\,Ivanov, {\it N=4 mechanics of general (4,4,0)
multiplets}, Nucl. Phys. B {\bf 855} (2012) 815-853, doi:10.1016/j.nuclphysb.2011.10.016,
{\tt arXiv:1107.1429 [hep-th]}.

\bibitem{HKT}
S.A.\,Fedoruk, E.A.\,Ivanov, A.V.\,Smilga, {\it N=4 mechanics with diverse (4,4,0) multiplets:
Explicit examples of HKT, CKT, and OKT geometries}, J. Math. Phys. {\bf 55} (2014) 052302, doi:10.1063/1.4871440,
{\tt arXiv:1309.7253 [hep-th]}.

\bibitem{QK} J.\,Bagger, E.\,Witten, {\it Matter couplings in N=2 supergravity}, Nucl. Phys. B {\bf 222} (1983) 1-10, doi:10.1016/0550-3213(83)90605-3.

\bibitem{MatN2}
A.\,Galperin, E.\,Ivanov, V.\,Ogievetsky, E.\,Sokatchev, {\it N=2 supergravity in superspace: Different versions and matter couplings},
Class. Quant. Grav. {\bf 4} (1987) 1255-1265, doi:10.1088/0264-9381/4/5/023.

\bibitem{GBIO}
J.\,Bagger, A.S.\,Galperin, E.A.\,Ivanov, V.I.\,Ogievetsky, {\it Gauging N=2 $\sigma$ models in harmonic superspace}, Nucl. Phys. B {\bf 303} (1988) 522-542,
doi:10.1016/0550-3213(88)90392-6.

\bibitem{GIOO}
A.\,Galperin, E.\,Ivanov, O.\,Ogievetsky, {\it Harmonic space and quaternionic manifolds}, Ann. Phys. {\bf 230} (1994) 201-249,
doi:10.1006/aphy.1994.1025, {\tt arXiv:hep-th/9212155}.


\bibitem{Ivanov:1999vg}
  E.~Ivanov and G.~Valent, {\it Quaternionic metrics from harmonic superspace: Lagrangian approach and quotient construction},
  Nucl.\ Phys.\ B {\bf 576} (2000) 543-577, doi:10.1016/S0550-3213(00)00178-4, {\tt arXiv:hep-th/0001165}.

\bibitem{ILe}
E.~Ivanov, O.~Lechtenfeld, {\it ${\cal N}=4$ supersymmetric mechanics in harmonic superspace}, JHEP {\bf 0309} (2003) 073, doi:10.1088/1126-6708/2003/09/073,
{\tt arXiv:hep-th/0307111}.

\bibitem{BISu}
S.~Bellucci, E.~Ivanov, A.~Sutulin, {\it ${\cal N}=8$ mechanics in $SU(2) \times SU(2)$ harmonic superspace}, Nucl. Phys. B {\bf 722} (2005) 297-327
[Erratum: Nucl. Phys. B {\bf 747} (2006) 464-465], doi:10.1016/j.nuclphysb.2005.06.018, 10.1016/j.nuclphysb.2006.04.017, {\tt arXiv:hep-th/0504185}.

\bibitem{DelSo}
F.~Delduc, E.~Sokatchev, {\it Superparticle with extended worldline supersymmetry}, Class. Quant. Grav. {\bf 9} (1992) 361-376, doi:10.1088/0264-9381/9/2/004.

\bibitem{PaSo}
A.I.~Pashnev, D.P.~Sorokin, {\it $n=4$ superfield description of relativistic spinning particle mechanics}, Phys. Lett. B {\bf 253} (1991) 301-305,
doi:10.1016/0370-2693(91)91723-9.

\bibitem{GNPW}
M.~G\"unaydin, A.~Neitzke, B.~Pioline, A.~Waldron, {\it Quantum Attractor Flows}, JHEP {\bf 0709} (2007) 056, doi:10.1088/1126-6708/2007/09/056, {\tt arXiv:0707.0267 [hep-th]}.

\bibitem{CLW}
D.~Cherney, E.~Latini, A.~Waldron, {\it Quaternionic K\"ahler Detour Complexes} \& {\it ${\cal N}=2$ Supersymmetric Black Holes}, Commun. Math. Phys. {\bf 302} (2011) 843-873,
doi:10.1007/s00220-010-1169-6, {\tt arXiv:1003.2234 [hep-th]}.

\bibitem{VanHolt}
M.P.~Bogers, J.W.~van Holten, {\it Canonical $D=1$ supergravity framework for FLRW cosmology}, JCAP 1505 (2015) no.05, 039, doi:10.1088/1475-7516/2015/05/039, {\tt arXiv:1503.01929 [gr-qc]}.

\bibitem{GaOO}
A.\,Galperin, O.\,Ogievetsky, {\it Harmonic potentials for quaternionic symmetric sigma models}, Phys. Lett. B {\bf 301} (1993) 67-71, doi:10.1016/0370-2693(93)90722-T,
{\tt arXiv:hep-th/9210153}.

\bibitem{HamAnalog}
A.\,Galperin, V.\,Ogievetsky, {\it N=2 D = 4 supersymmetric sigma models and Hamiltonian mechanics}, Class. Quant. Grav. {\bf 8} (1991) 1757-1764, doi:10.1088/0264-9381/8/10/004.


\bibitem{Gun}
M. \, G\"unaydin, {\it Harmonic Superspace, Minimal Unitary Representations and Quasiconformal Groups}, JHEP {\bf 0705} (2007) 049,
doi:10.1088/1126-6708/2007/05/049, {\tt arXiv:hep-th/0702046};\\

M. \, G\"unaydin, {\it Lectures on Spectrum Generating Symmetries and U-duality in Supergravity, Extremal Black Holes, Quantum Attractors and Harmonic Superspace},
Springer Proc. Phys. {\bf 134} (2010) 31-84, doi:10.1007/978-3-642-10736-8-2,
{\tt arXiv:0908.0374 [hep-th]}.

\bibitem{HOP}
P.S.\,Howe, A.\,Opfermann, G.\,Papadopoulos, {\it Twistor spaces for QKT manifolds}, Commun. Math. Phys. {\bf 197} (1998) 713-727,
doi:10.1007/s002200050469, {\tt arXiv:hep-th/9710072}.

\bibitem{IvNied}
E.~Ivanov, J.~Niederle, {\it Bi-Harmonic Superspace for N=4 Mechanics}, Phys. Rev. D {\bf 80} (2009) 065027,
doi:10.1103/PhysRevD.80.065027, {\tt arXiv:0905.3770 [hep-th]}.

\bibitem{ReviewSC}
S.~Fedoruk, E.~Ivanov, O.~Lechtenfeld, {\it Superconformal mechanics}, \\
J. Phys. A {\bf 45} (2012) 173001, doi:10.1088/1751-8113/45/17/173001, {\tt arXiv:1112.1947 [hep-th]}.

\end{thebibliography}
\end{document}